\documentclass[aps,pre,superscriptaddress,twocolumn,10pt]{revtex4-1}
\usepackage[T1]{fontenc}
\pdfoutput=1
\usepackage[pdftex]{graphicx}
\usepackage{lmodern}
\usepackage[usenames,dvipsnames,cmyk]{xcolor}
\usepackage{amsmath,amsfonts}
\usepackage[english]{babel}
\usepackage[pdftex,hyperfigures,breaklinks,colorlinks,pdfusetitle]{hyperref}
\definecolor{LinkColor}{rgb}{0, 0, 0.5}
\definecolor{ExtLinkColor}{rgb}{0, 0.4, 0}
\hypersetup{citecolor=LinkColor,linkcolor=LinkColor,urlcolor=ExtLinkColor,filecolor=ExtLinkColor}
\renewcommand{\vec}[1]{\mathbf{#1}}
\newcommand{\avg}[1]{\langle #1 \rangle}

\renewcommand{\d}[1]{\mathrm{d}#1}
\newcommand{\dd}[1]{\mathrm{d}^{2}\hspace{-.1em}#1}
\newcommand{\I}{\mathrm{i}}
\newcommand{\E}{\mathrm{e}}

\DeclareMathOperator{\Tr}{Tr}

\renewcommand{\Im}{\operatorname{Im}}
\renewcommand{\Re}{\operatorname{Re}}

\begin{document}

\title{Correlations between synapses in pairs of neurons slow down dynamics in randomly connected neural networks}

\author{Daniel Mart\'{\i}}
\affiliation{Laboratoire de Neurosciences Cognitives,\\
Inserm UMR No.~960, Ecole Normale Sup\'{e}rieure,\\
PSL Research University, Paris, France}
\author{Nicolas Brunel}
\affiliation{Department of Statistics and Department of Neurobiology,\\
University of Chicago, Chicago, IL 60637, USA}
\affiliation{Department of Neurobiology and Department of Physics,\\
Duke University, Durham, NC 27710, USA}
\author{Srdjan Ostojic}
\affiliation{Laboratoire de Neurosciences Cognitives,\\
Inserm UMR No.~960, Ecole Normale Sup\'{e}rieure,\\
PSL Research University, Paris, France}

\date{\today}

\begin{abstract}
 Networks of randomly connected neurons are among the most popular models in
 theoretical neuroscience. The connectivity between neurons
 in the cortex is however not fully random, the simplest and
 most prominent deviation from randomness found in experimental data
 being the overrepresentation of bidirectional connections among
 pyramidal cells. Using numerical and analytical methods, we
 investigated the effects of partially symmetric connectivity on
 the dynamics in networks of rate units. We consider the two dynamical
 regimes exhibited by random neural networks: the weak-coupling
 regime, where the firing activity decays to a single fixed point
 unless the network is stimulated, and the strong-coupling or chaotic
 regime, characterized by internally generated fluctuating firing
 rates. In the weak-coupling regime, we compute analytically, for an
 arbitrary degree of symmetry, the autocorrelation of network
 activity in the presence of external noise. In the chaotic regime, we
 perform simulations to determine the timescale of the intrinsic
 fluctuations. In both cases, symmetry increases the characteristic
 asymptotic decay time of the autocorrelation function and therefore
 slows down the dynamics in the network.
\end{abstract}

\maketitle

 \section*{Introduction}

 The dynamics and function of a network of neurons is to a large
 extent determined by its pattern of synaptic connections. In the
 mammalian brain, cortical networks exhibit a complex connectivity
 that to a first approximation can be regarded as random. This
 connectivity structure has motivated the study of networks of
 neurons connected through a random synaptic weight matrix with
 independent and identically distributed (i.i.d.) entries, which have become a
 central paradigm in theoretical
 neuroscience~\citep{sompolinsky_crisanti_sommers_prl1988,vanvreeswijk_sompolinsky_neco1998,brunel_jcns2000}.
 Randomly connected networks of firing-rate units exhibit a chaotic phase~\citep{sompolinsky_crisanti_sommers_prl1988}, which can be exploited as a
 susbstrate for complex
 computations~\citep{buonomano_maass_natns2009,toyoizumi_abbott_pre2011,sussillo_abbott_neuron2009}.
 Networks of randomly connected spiking neurons also exhibit rich dynamics that can account for
 the highly irregular spontaneous activity observed in
 the cortex \emph{in
 vivo}~\citep{shadlen_newsome_jns1998,vanvreeswijk_sompolinsky_neco1998,amit_brunel_cercor1997,brunel_jcns2000}.
 Importantly, these models are to a large extent amenable to a mathematical
 analysis, which allows for a thorough understanding of the
 mechanisms underlying their dynamics.

 Detailed analyses of experimental data on cortical connectivity have however
 identified patterns of connectivity that strongly deviate from the
 i.i.d.~assumption~\citep{markram_etal_jphysiol1997,song_etal_plosbiol2005,perin_etal_pnas2011,ko_etal_nature2011,harris_mrsic-flogel_nature2013}. The
 most prominent of such deviations is the overrepresentation of
 reciprocal
 connections~\citep{markram_etal_jphysiol1997,song_etal_plosbiol2005,wang_etal_natns2006},
 and the fact that synapses of bidirectionally connected pairs of
 neurons are on average stronger than synapses of unidirectionally
 connected pairs. These observations are consistent with a partially
 symmetric connectivity structure, intermediate between full symmetry
 and full asymmetry. How partial symmetry in the connectivity impacts
 network dynamics is not yet understood, in part because such
 partial symmetry renders the mathematical analyses more
 challenging~\citep{crisanti_sompolinsky_pra1987}. Here we study the
 impact of partial symmetry in the connectivity structure on the dynamics
 of a simple network model consisting of interacting rate
 units. Depending on the overall strength of coupling, such a network
 can display either a stable or a chaotic regime of activity, as in
 the random asymmetric case~\citep{sompolinsky_crisanti_sommers_prl1988}. We examined how the
 degree of symmetry in the network influences the temporal dynamics
 in both regimes. For the stable regime, we exploited recent results
 from random matrix theory~\citep{chalker_mehlig_prl1998,mehlig_chalker_jmathphys2000} to
 derive analytical expressions for the autocorrelation
 functions. These expressions demonstrate that increasing the
 symmetry in the network leads to a slowing down of the
 dynamics. Numerical simulations in the chaotic regime show a similar
 effect, with time scales increasing far more substantially with
 symmetry than in the fixed point regime. Altogether, our results indicate that symmetry in the connectivity can act as an additionnal source of slow dynamics, an important ingredient for implementing computations in networks of neurons \citep{huang_doiron_coinb2017}.

 \section{Description of the model}

 We consider a network of $N$ fully connected neurons, each described 
 by an activation variable (synaptic current) $x_i$, $i=1,\dotsc,N$, obeying
 \begin{equation}
 \frac{\d{x}_i}{\d{t}} = -x_i + g \sum_{j=1}^{N}
 J_{ij} \phi(x_j), \label{eq:circuit}
 \end{equation}
 where $g$ is a gain parameter that modulates the strength of
 recurrent connections, and where $\phi(\cdot)$ is the input-output
 transfer function that transforms activations $x_i$ into firing
 rates. This transformation is non-linear and we model it as
 $\phi(x)=\tanh(x)$ for mathematical convenience (see~\citep{rajan_abbott_sompolinsky_pre2010,kadmon_sompolinsky_prx2015,harish_hansel_ploscb2015,mastrogiuseppe_ostojic_2016} for studies of network models with
 different choices of $\phi$). The elements $J_{ij}$ of the
 connectivity matrix are drawn from a Gaussian distribution with zero
 mean, variance $1 / N$, and correlation
 \[ [J_{ij} J_{ji}]_J = \eta / N,\]
 with the square brackets $[\cdot]_{J}$ denoting an average over
 realizations of the random connections. The parameter $\eta$ is the
 correlation coefficient between the two weights connecting pairs of
 neurons, and quantifies the degree of symmetry of the
 connections. For $\eta=0$ the elements $J_{ij}$ and $J_{ji}$ are
 independent and the connectivity matrix is fully asymmetric; for
 $\eta=1$ the connectivity matrix is fully symmetric; For $\eta=-1$ it
 is fully antisymmetric.  In Secs.~I--IIIA we study the full range
 $\eta\in[-1,1]$, while in Secs~IIIB--IV we focus on $\eta\in[0,1]$.

 \section{Dynamical regimes of the network}

 For fully asymmetric matrices, previous work has shown that the
 network activity described by \eqref{eq:circuit} undergoes a phase
 transition at $g=1$ in the limit of large $N$~\citep{sompolinsky_crisanti_sommers_prl1988}. For $g < 1$ the
 activity for all units decays to 0, which is the unique stable fixed point of the
 dynamics \citep{wainrib_touboul_prl2013}, while for $g > 1$ the activity is chaotic. Such a
 transition can be partially understood by assessing the stability of
 the fixed point at $x_i=0$ for $i=1, \dotsc, N$. If we linearize
 Eq.~\eqref{eq:circuit} around this fixed point we obtain the
 stability matrix, with components
 \begin{equation}
 M_{ij} = -\delta_{ij} + gJ_{ij}.
 \label{eq:stability_mat}
 \end{equation}
 The eigenvalues of $M_{ij}$ are therefore those of the matrix $J_{ij}$, scaled
 by the gain $g$ and shifted along the real axis by $-1$. In the limit
 $N\rightarrow \infty$, for a connectivity matrix $J_{ij}$ whose entries are
 i.i.d.~Gaussian random variables of zero mean and variance $1/N$, eigenvalues
 are uniformly distributed in the unit disk of the complex
 plane~\citep{ginibre_jmatphys1965,girko_thappprob1984,tao_vu_annprob2010}.
 This implies that the eigenvalues of the stability matrix have a negative real
 part as long as $g < 1$, and therefore that the fixed point at 0 is stable in
 that range.

\begin{figure*}[htbp]
 \centering
 \includegraphics{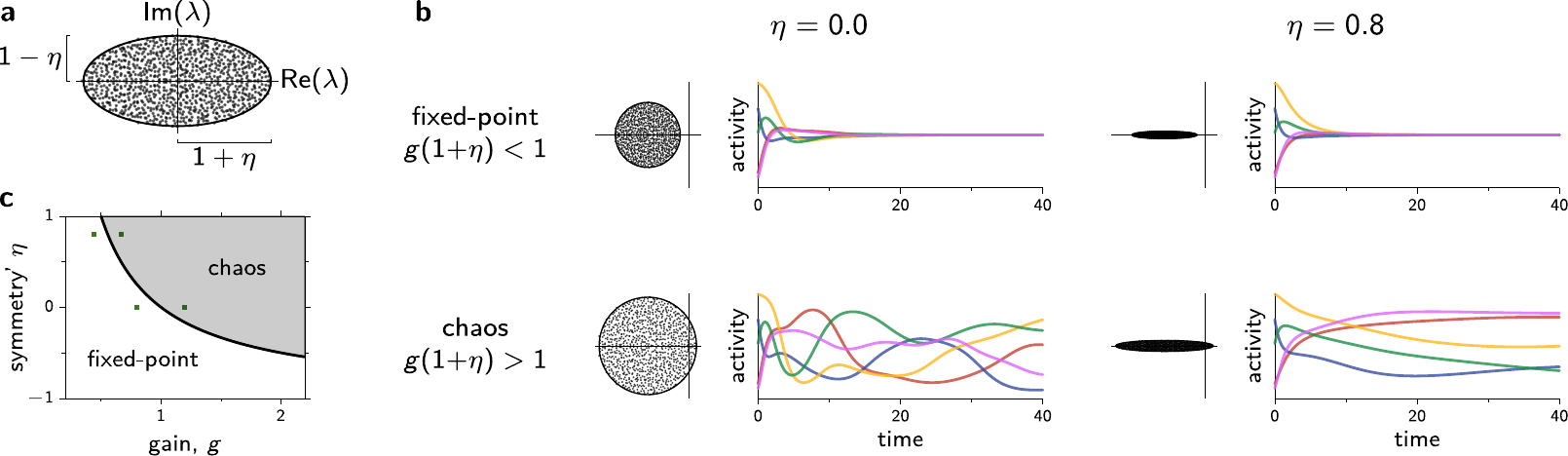}
 \caption{
 \label{fig:one}
   \textbf{a} Spectrum of eigenvalues of a Gaussian random
 matrix with zero mean, variance $1/N$, and correlation coefficient $\eta$
 between weights connecting neuronal pairs. \textbf{b} Time evolution of the
 firing rates of five arbitrary
 neurons, for the fixed-point regime (top) and the chaotic regime
 (bottom), and for two different values of $\eta$ (left and right panels).
 Next to each trajectory is the eigenspectrum of the corresponding
 linearized system along with the real and imaginary axes, which we include
 for reference. The initial firing rates were the same for all panels. All
 connectivity matrices were derived from a single realization of a Gaussian
 random matrix. To generate connectivity matrices with different $\eta$, we
 used the upper right and the lower left triangular portions of the Gaussian
 random matrix to create a symmetric and antisymmetric matrix, which we then
 combined to yield a $\mathbf{J}$ with the desired symmetry parameter $\eta$
 [for details, see the text surrounding Eq.\eqref{eq:partial_sym_build}, in
 Appendix~\ref{sec:app_DMF}]. \textbf{c} Activity regimes as a function of
 the gain and the degree of symmetry in the connections. The green squares
 indicate the parameter values used in \textbf{b}.}
 \end{figure*}

 An analogous transition occurs when connections are partially
 symmetric. The presence of correlations among weights deforms the
 spectrum of eigenvalues into an ellipse, elongating its major radius
 by a factor of $1 + \eta$ and shortening the minor radius by a
 factor $1 - \eta$ \citep{girko_teorver1985,sommers_etal_prl1988,naumov_arxiv2012,nguyen_orourke_imrn2014}
 [Fig.~\ref{fig:one}a]. This property is usually referred to as the
 elliptic law. For the network described by~\eqref{eq:circuit} such a
 deformation causes the fixed point at $x_i = 0$ for $i=1,\dotsc,N$
 to lose its stability at $g = 1 / (1 + \eta)$
 [Figs~\ref{fig:one}b and \ref{fig:one}c]. In other words, symmetry lowers the
 critical coupling.


 Our goal is to characterize how the degree of symmetry in the
 connections affects the network activity on each side of the
 instability: the relaxation response of the network at low gains and
 the chaotic self-generated activity observed at strong
 gains. Our description of the network activity will be based on the
 average autocorrelation function,
 \begin{equation}
 C(\tau) = \frac{1}{N} \sum_{i=1}^{N} \bigl[\avg{x_i(t) x_i(t + \tau)}\bigr]_J,
 \label{eq:C_estimate}
 \end{equation}
 where the average is over both the population and the realizations of the
 connectivity matrix \citep{sompolinsky_crisanti_sommers_prl1988}, and where
 we are assuming for now that the system is stationary.

\section{Dynamics in the fixed-point regime}

\subsection{Derivation of the autocorrelation function}
\label{sec:derivation_ac}

\begin{figure*}
 \includegraphics{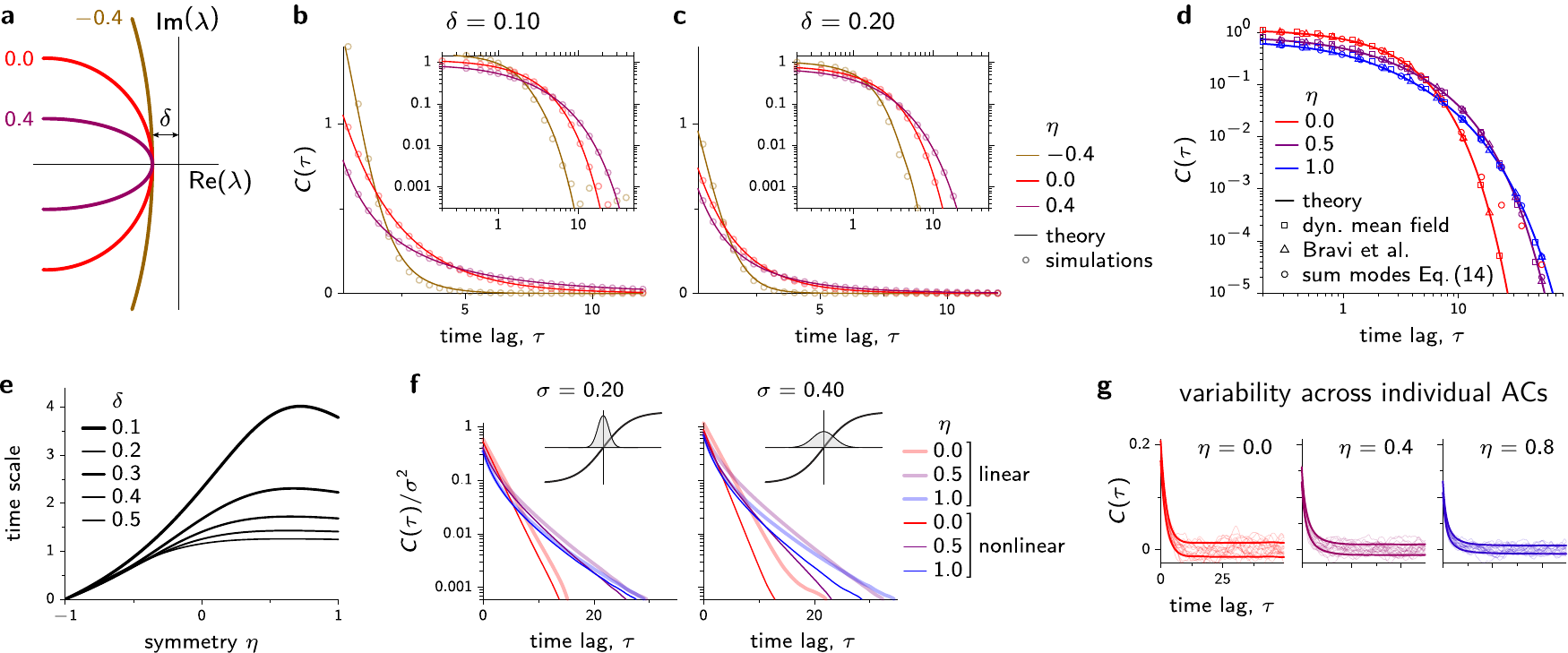}
\caption{
\label{fig:two}
  \textbf{a} The distance of the eigenspectrum to the imaginary axis, or
spectral gap, is kept fixed at a value $\delta$ independently of the symmetry
parameter $\eta$. The color curves represent the rightmost portion of the boundaries
of the eigenspectra for $\eta=-0.4, 0, 0.4$ (values indicated next to
each curve). \textbf{b},\textbf{c} Analytical prediction and numerical estimate
of the average autocorrelation, for different values of the symmetry parameter
$\eta$, indicated in the key. Each subplot corresponds to a particular spectral
gap $\delta$. Insets show the same curves on a log-log scale. The numerical
estimate of $C(\tau)$ was derived from Eq.~\eqref{eq:C_estimate}, using
simulated neuronal activity generated by Eq.~\eqref{eq:linear}, with $N=10000$
units, and averaging over time, units, and 200 different realizations of the
connectivity matrix. \textbf{d} Comparison of our analytical prediction with
three alternative semi-numerical predictions (see the text for details). \textbf{e} Dependence on \(\eta\) of the timescale \(\hat{\tau}\), estimated here as the mean of an unnormalized distribution defined by \(C(\tau)\): \(\hat{\tau} = \int_{0}^{\infty} s C(s)\,\d{s} /
\int_{0}^{\infty} C(s)\,\d{s}\). \textbf{f} Changes in the autocorrelation function induced by the nonlinear activation function \(\Phi(x)=\tanh(x)\), for two values of the amplitude of the injected noise. The inset shows noise distribution superimposed on \(\Phi(x)\). Autocorrelations were numerically estimated as in \textbf{b}, \textbf{c}. \textbf{g} Samples of 20 individual autocorrelation functions for three different values of \(\eta\). Solid thick curves indicate the bounds of the interval $[\text{mean} - \text{std}, \text{mean} + \text{std}]$, where the standard deviation was estimated from the full population of of individual autocorrelations.}
\end{figure*}

In the fixed-point regime, the activity decays to zero unless the
network is stimulated by external inputs. To characterize the dynamics
of the network in this regime, we induce network activity by feeding
each neuron with independent Gaussian white noise \cite{schuecker_goedeke_helias_arxiv2016}. The amplitude of
this noise is assumed to be small enough so that the synaptic
activation of all neurons lies within the linear range of their
input-to-rate transfer function [see Fig.~\ref{fig:two}f for the range of validity of that approximation]. Under these conditions, $\phi(x)$ can
be approximated by its first order Taylor expansion $\phi(0) +
\phi'(x)|_{x=0} x = x$, and the dynamical equations become
 \begin{equation}
 \frac{\d{\vec{x}(t)}}{\d{t}} = (-\mathbf{1} + g\mathbf{J})\vec{x}(t) + \sigma \boldsymbol{\xi}(t),
 \label{eq:linear}
 \end{equation}
 where $\vec{x}(t)=(x_1(t), \dotsc, x_N(t))^{T}$, $\mathbf{1}$ is the
 identity matrix, $\vec{J}$ is the connectivity matrix, and
 $\boldsymbol{\xi}(t) = (\xi_1(t), \dotsc, \xi_N(t))^{T}$ is a vector
 of independent white noise sources of zero mean and unit variance:
 $\avg{\xi_i(t)} = 0$, $\avg{\xi_i(t) \xi_j(t')} = \delta_{ij}
 \delta(t - t')$, with angular brackets representing averages over
 noise realizations. The parameter $\sigma$ is the standard deviation
 of the white noise injected into neurons.

 The time scales displayed by a linear system like~\eqref{eq:linear}
 are strongly affected by the real part of the eigenvalues of the
 system's stability matrix and, in particular, they get longer as
 eigenvalues get closer to the imaginary axis. To disentangle this
 type of slowing down from the effects due to symmetry alone, we vary
 the parameter $\eta$ while keeping the \emph{spectral gap} fixed. By
 spectral gap we mean the distance between the spectrum of
 eigenvalues of the stability matrix $M_{ij}$,
 Eq.~\eqref{eq:stability_mat}, and the imaginary axis [Fig.~\ref{fig:two}a]. From the elliptic law, the eigenvalue of the
 stability matrix with the largest real part is $z = -1 + g(1 + \eta)$,
 and we can keep the spectral gap at $\delta$ by setting the gain to
 $g = (1 - \delta) / (1 + \eta)$.

The system described by \eqref{eq:linear} is linear and can be solved
by diagonalizing the connectivity matrix. The matrix $\vec{J}$ admits
a set of right eigenvectors $\{\vec{R}_1, \dotsc, \vec{R}_N\}$ that
obey $\vec{J} \vec{R}_i = \lambda_i \vec{R}_i$ for
$i=1,\dotsc,N$. These eigenvectors are in general complex-valued and,
except for the symmetric case $\eta=1$, not orthogonal to one another,
which implies that $\vec{J}$ cannot be diagonalized through a unitary
transformation. Matrices of this kind are called non-normal and
do not commute with their transpose conjugate: $\vec{J}\vec{J}^{\dag}
\neq \vec{J}^{\dag}\!\vec{J}$~\citep{trefethen_embree_2005}. Even if
non-normal matrices cannot be diagonalized by an orthogonal set of
eigenvectors, it is always possible to form a biorthogonal
basis by extending the set of right eigenvectors with the set of left
eigenvectors, which obey $\vec{L}_i^{\!\dag} \vec{J} = \lambda_i
\vec{L}^{\!\dag}_i$. This extended basis is biorthogonal in the sense
that $\vec{L}^{\dag}_i \vec{R}_j = \delta_{ij}$. We can summarize all
these properties in a compact way by defining the square matrices
$\vec{R}$ and $\vec{L}$ that result from adjoining in columns the set
of, respectively, right and left eigenvectors, and by introducing the
diagonal matrix $\boldsymbol{\Lambda}$ that contains the eigenvalues
$\lambda_i$ of $\vec{J}$ in its diagonal entries. In this notation the
biorthogonality condition is $\vec{L}^{\!\dag}\vec{R} = \vec{1}$ and
the eigenvalue equations for the right and left eigenvectors read
$\vec{J}\vec{R} = \vec{R}\boldsymbol{\Lambda}$ and $\vec{L}^{\!\dag}\vec{J} =
\boldsymbol{\Lambda}\vec{L}^{\!\dag}$.

We can now write the formal solution of~\eqref{eq:linear}:
\begin{equation*}
 \begin{split}
 \vec{x}(t) & = \sigma \int_{-\infty}^{t} \E^{(-\vec{1} + g\vec{J})(t -
s)}\boldsymbol{\xi}(s)\,\d{s}\\
& = \sigma \vec{R} \int_{-\infty}^{t} \E^{(-\vec{1} + g \boldsymbol{\Lambda})(t -
s)} \vec{R}^{-1}\boldsymbol{\xi}(s)\,\d{s}\,,
\end{split}
\end{equation*}
where in the last equality we used the basis of right eigenvectors to
write $\vec{J} = \vec{R} \boldsymbol{\Lambda}\vec{R}^{-1}$ and we
implicitly expanded the exponential in its power series to obtain the
final result. From this expression we can derive the
population-average autocorrelation for a particular realization of the
connectivity:
\begin{align}
 C_{J}(\tau) & = \frac{1}{N} \avg{ \vec{x}^{\dag}\!(t)
 \vec{x}(t + \tau)} =
 \frac{1}{N} \Tr \avg{\vec{x}(t + \tau)
 \vec{x}^{\dag}\!(t)}\notag \\ 
 & = \frac{\sigma^2}{N} \! \int_{0}^{\infty}
 \hspace{-0.5ex} \E^{-2 u - \tau} \Tr
 \bigl\{ \vec{R}^{\!\dag}\vec{R}\,\E^{g\boldsymbol{\Lambda}(u + \tau)}
 \vec{L}^{\!\dag}\vec{L}\, \E^{g\boldsymbol{\Lambda}^{\!\dag}u}\bigr\} \,\d{u}.
 \label{eq:Ctau_sums}
\end{align}
In the second equality we used the cyclicity of the trace, and in the
last line we changed the integration variable to $u=t - s$ and we used
the biorthogonality condition to write $\vec{R}^{-1} =
\vec{L}^{\!\dag}$. The average over noise amounts to applying the
identity $\langle \boldsymbol{\xi}(t)\boldsymbol{\xi}^{\dag}\!(t')
\rangle = \sigma^2 \mathbf{1} \delta(t - t')$. Note that the
$\sigma^2$ appears as an overall factor, so we can set $\sigma=1$
without loss of generality.

We can simplify~\eqref{eq:Ctau_sums} by introducing the so-called overlap matrix, with components
\begin{equation}
 O_{ij} = (\vec{L}^{\!\dag}\vec{L})_{ij} (\vec{R}^{\!\dag}\vec{R})_{ji},
 \label{eq:overlap}
\end{equation}
and which characterizes the correlations between left and right eigenvectors~\citep{chalker_mehlig_prl1998}. Equation~\eqref{eq:Ctau_sums} then becomes
\begin{equation}
 C_{J}(\tau) = \frac{1}{N} \int_{0}^{\infty} \hspace{-1ex}
 \E^{-2 u - \tau}\sum_{i=1}^{N}\sum_{j=1}^{N} \E^{g \lambda_i(u + \tau)}
 O_{ij} \E^{g \bar{\lambda}_{j}u}\,\d{u}.
 \label{eq:finite_dim_overlap}
\end{equation}
If the connectivity matrix were normal, the overlap would be the
identity matrix and the autocorrelation~\eqref{eq:finite_dim_overlap}
would just be a sum of independent contributions---one per
eigenvalue. These contributions are coupled for non-normal matrices.

We can make further analytical progress by studying the
autocorrelation~\eqref{eq:finite_dim_overlap} in the limit $N
\rightarrow \infty$, in which the differences of $C_J(\tau)$ across
realizations of the connectivity matrix disappear, and the autocorrelations of all units become close to the population average [Fig.~\ref{fig:two}g]. In that limit sums
over indices are replaced with integrals over eigenvalues, while the
overlap matrix is replaced with the local average of the overlap,
defined as
\begin{equation}
 D(z_1, z_2) = \!
 \lim_{N\rightarrow \infty} \frac{4}{N} \left[ \sum_{i=1}^{N} \sum_{j=1}^{N} O_{ij}
 \delta^2(z_1 - \lambda_{i}) \delta^2(z_2 -
\lambda_{j})\right]_{\!J}\!\!\!.
\label{eq:densityD}
\end{equation}
Here $z=x + \I y$ are complex numbers and we defined the complex Dirac
delta as $\delta^{2}(z) \equiv (1/2) \delta(x) \delta(y)$ so that it
satisfies the normalization condition $\int \delta^{2}(z)\, \dd{z}
\equiv \int \delta^{2}(z)\, \d{z}\d{\bar{z}} = 1$. After taking the
limit $N\rightarrow \infty$, Eq.~\eqref{eq:finite_dim_overlap}
becomes
\begin{equation}
 C(\tau) = \int_{0}^{\infty} \E^{-2 u - \tau} A(u, \tau)\,\d{u}.
 \label{eq:Ctau_Adef}
\end{equation}
where we defined
\begin{equation}
 A(u, \tau) = \frac{1}{4} \iint \E^{g(z_1 + \bar{z}_2) u + g z_1 \tau} D(z_1,
 z_2)\,\dd{z_1} \dd{z_2}.
 \label{eq:Adef}
\end{equation}
Each of the integrals in Eq.~\eqref{eq:Adef} is over complex values,
and involves the expression of $D(z_1, z_2)$ for the ensemble of
Gaussian random matrices with partial symmetry, which was derived
using diagrammatic techniques in \citep{chalker_mehlig_prl1998} and
whose functional form can be found in Appendix~\ref{sec:app_CM}. We
used the result of \citep{chalker_mehlig_prl1998} to evaluate the
double complex integral $A(u, \tau)$ in Eq.~\eqref{eq:Adef}. The
details of the evaluation are given in Appendix~\ref{sec:app_CM}, and
the result is
\begin{equation}
 A(u, \tau) = A_1(u, \tau) + A_2(u, \tau),
 \label{eq:Adef_decomposition}
\end{equation}
with
\begin{align}
 A_1(u, \tau) & = (1 + \eta^2) I_0\bigl(g\, \psi(u, \tau; \eta) \bigr)\notag \\ 
 & - 2 \eta \left(1 + \frac{2(1 -\eta)^2\tau^2}{\psi(u, \tau; \eta)^2} \right)
 I_2\bigl(g \,\psi(u, \tau; \eta)\bigr), \label{eq:A1}\\
 A_2(u, \tau) & = \frac{-1}{g^2 u (u + \tau)} \sum_{k=1}^{\infty} \eta^{k} k^2 I_{k} \bigl(2g\sqrt{\eta}u\bigr)\notag\\
 & \hspace{9em} \times I_{k} \bigl(2g\sqrt{\eta}(u + \tau)\bigr), \label{eq:A2}
\end{align}
where $I_k(\cdot)$ is the modified Bessel function of order $k$, and
where in Eq.~\eqref{eq:A1} we defined
\[\psi(u, \tau;\eta) = 2 \sqrt{(1 + \eta)^2 u (u + \tau) + \eta \tau^2}.\]
The autocorrelation is finally computed from Eq.~\eqref{eq:Ctau_Adef},
integrating numerically over $u$.

Expressions.~\eqref{eq:Adef_decomposition}, \eqref{eq:A1},
\eqref{eq:A2} are valid for full range \(-1 \leq \eta \leq 1\). For negative
\(\eta\) we replace \(\sqrt{\eta}\) with \(\I\sqrt{|\eta|}\) and apply the
identity \(I_{\nu}(\I z) = \I^{\nu} J_\nu(z)\), which is valid for integer
\(\nu\).

The analytical prediction given by Eqs.~\eqref{eq:Ctau_Adef} and
Eqs.~\eqref{eq:Adef_decomposition}--\eqref{eq:A2} matches with the
autocorrelation estimated from numerical simulations
[Fig.~\ref{fig:two}b], although for long time lags the numerical
estimate becomes noisy due to finite-size effects. To check the
validity of our prediction also at long time lags, we compared our
analytical prediction with three alternative derivations
[Fig.~\ref{fig:two}c]. One such derivation consists of estimating
the autocorrelation for large but finite $N$, by computing numerically
the eigenvalues and eigenvectors of randomly generated matrices,
evaluating the time integral of Eq.~\eqref{eq:finite_dim_overlap},
which gives
\begin{equation}
 C_J(\tau) = -\frac{1}{N} \sum_{i=1}^{N} \sum_{j=1}^{N} \frac{O_{ij} \E^{-(1 - g\lambda_i) \tau}}{2 + g(\lambda_i + \bar{\lambda}_j)},
 \label{eq:sum_modes}
\end{equation}
and then by averaging $C_J(\tau)$ over multiple realizations of the
connectivity matrix. Another derivation is based on dynamical
mean-field
theory~\citep{sompolinsky_zippelius_prb1982,crisanti_sompolinsky_pra1987,schuecker_goedeke_helias_arxiv2016},
which gives rise to a set of integro-differential equations involving
$C(\tau)$ that can be solved numerically
(Appendix~\ref{sec:app_DMF}). Finally, we numerically computed the
inverse Fourier transform of the power spectrum derived in
\citep{bravi_sollich_opper_jpa2016} for this same system.
\citet{bravi_sollich_opper_jpa2016} used a perturbative method to
derive the system of integro-differential
equations~\eqref{eq:SD1lin} and \eqref{eq:SD2lin}, which they solved for
the correlation and reponse functions by using a Laplace transform.
All derivations yield the same result, except for the deviations we
observe when applying Eq.~\eqref{eq:sum_modes} at long $\tau$,
which are caused by finite-size effects.

Our results show that an increase in symmetry tends to spread
autocorrelations toward longer time lags, and that this effect gets
larger the closer the system gets to the onset of chaos [Fig.~\ref{fig:two}e]. An intuitive
explanation for this slowing down is that the deformation of the
eigenspectrum caused by symmetry increases the density of eigenvalues
with small imaginary parts, thereby enlarging the contribution of
low-frequency modes.

\subsection{Behavior at long time lags}

\begin{figure*}
 \centering
 \includegraphics{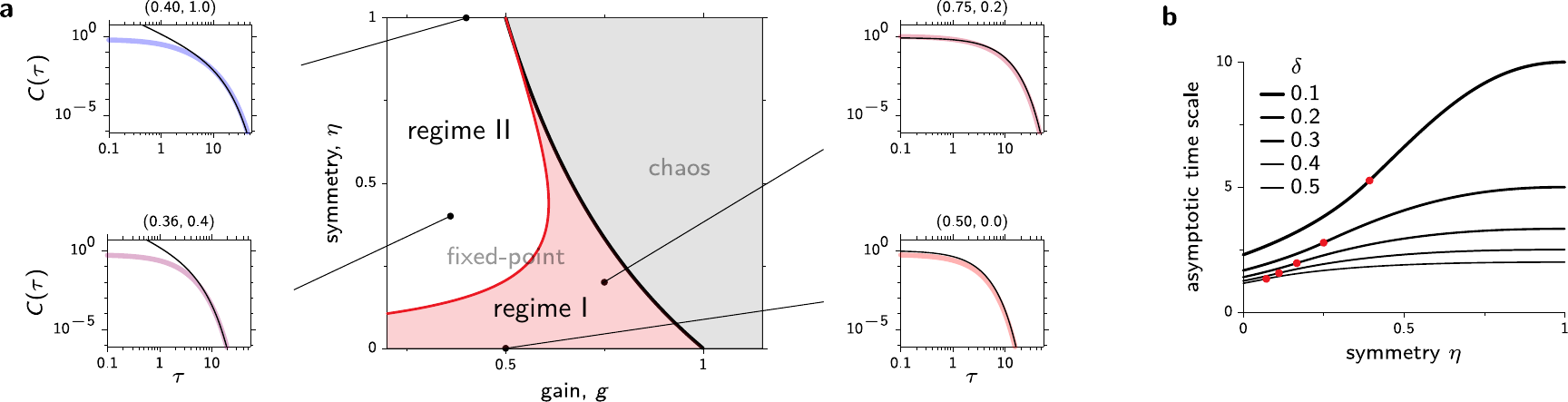}
 \caption{\label{fig:three}
   \textbf{a} Phase diagram showing the different activity regimes as a function of the gain $g$ and the degree of symmetry $\eta$ (center). At both sides of the diagram, we show the autocorrelation functions at a few representative points in the $(g,\eta)$ plane, indicated by the black dots in the main diagram. Each autocorrelation plot contains the exact prediction of $C(\tau)$ given by Eqs.~\eqref{eq:Ctau_Adef}--\eqref{eq:A2} (thick colored curve) and the asymptotic approximation, summarized in Eqs.~\eqref{eq:C_asymp_exp}--\eqref{eq:G_II} (thin black curve). At the top of each such plot are the parameters $g$ and $\eta$ used. \textbf{b} Asymptotic time scale of the autocorrelation as a function of the degree of symmetry, for different spectral gaps $\delta$. These curves were generated with the inverses of $G_{\text{I},\text{II}}(\eta,\delta)$, Eqs.~\eqref{eq:C_asymp_exp} and \eqref{eq:C_asymp_powerlaw}. The red dot on each curve indicates the value of $\eta$ where the transition between subregimes occurs.}
\end{figure*}
While equations~\eqref{eq:Ctau_Adef}--\eqref{eq:A2} are exact, they provide little analytical insight into how the autocorrelation depends on parameters. A more explicit dependence can be obtained by evaluating $C(\tau)$ in the limit of long $\tau$. We relegate the details of the calculation to Appendix~\ref{sec:app_laplace} and summarize the main results here. The analysis shows that, in the fixed point regime, there exist two subregimes of activity that differ in how the asymptotic decay rate of the autocorrelation depends on the symmetry parameter $\eta$ and the spectral gap $\delta$. For small values of $\eta$ and $\delta$, the autocorrelation decays as a pure exponential at long $\tau$ (regime I),
\begin{equation}
C(\tau) = F_{\text{I}}(\eta, \delta)\, \E^{-\tau G_{\text{I}}(\eta, \delta)},
 \label{eq:C_asymp_exp}
\end{equation}
with
\begin{align}
 F_{\text{I}}(\eta, \delta) & = \frac{\delta^{-1/2}(1 - \eta)^2}{2\sqrt{2} (1 - \delta)},\label{eq:F_I} \\
 G_{\text{I}}(\eta, \delta) & = \frac{1 - \eta}{1 + \eta} \sqrt{2\delta - \delta^2}.\label{eq:G_I}
\end{align}
Conversely, for sufficiently large values of $\eta$ and $\delta$ autocorrelation for long $\tau$ can be approximated by a power multiplied by an exponential decay (regime II):
\begin{equation}
 C(\tau) = \tau^{-3/2} F_{\text{II}}(\eta, \delta)\, \E^{-\tau G_{\text{II}}(\eta, \delta)},
 \label{eq:C_asymp_powerlaw}
\end{equation}
with
\begin{align}
 F_{\text{II}}(\eta, \delta) & = \frac{1}{4\sqrt{\pi}} \left( \frac{1 + \eta}{1 - \delta}\right)^{3/2} \biggl[ \frac{ 2 \eta^{-1/4}(1 + \eta^2)}{\delta(1 + \eta) - [1 - \sqrt{\eta}]^2} \phantom{aa}\notag\\
& \hspace*{2.8cm}{- \frac{\eta^{5/6} (1 + \eta)}{(1 - \sqrt{\eta})^2 + 2 \sqrt{\eta}\delta} \biggr]}\label{eq:F_II} \\
G_{\text{II}}(\eta, \delta) & = \frac{(1 - \sqrt{\eta})^2 + 2\delta \sqrt{\eta}}{1 + \eta}.\label{eq:G_II}
\end{align}
A comparison between the asymptotic expression in
Eq.~\eqref{eq:C_asymp_powerlaw} and the full expression for the
autocorrelation function reveals, however, that the power law is not
observed in practice because the range below the cutoff falls below
the values of $\tau$ where the asymptotic approximation starts
matching the exact expression.

Figure~\ref{fig:three}a shows the exact parameter region of each asymptotic regime, after transforming the spectral gaps into gains. In both regimes the autocorrelation's asymptotic decay rate matches the exact result for time lags longer than a few time units [see Fig.~\ref{fig:three}a, lateral panels]. It seems therefore reasonable to associate the time scale of the autocorrelation with the inverse of $G_{\text{I},\text{II}}(\eta,\delta)$ [see Eqs.~\eqref{eq:C_asymp_exp} and \eqref{eq:C_asymp_powerlaw}], where the subindex $\text{I}, \text{II}$ is chosen according to the subregime found at the parameter values $(\eta, \delta)$. The asymptotic time scale of the autocorrelation increases monotonically with symmetry regardless of the subregime the network operates in [Fig.~\ref{fig:three}b], although this dependence is convex in the exponential subregime and concave in the power-law-with-cutoff regime [in Fig.~\ref{fig:three}b see the curves split by the red dots, which mark the boundary between subregimes]. Note also that as the spectral gap $\delta$ shrinks to 0 the system enters the exponential regime and timescales diverge as $\delta^{-1/2}$, according to Eq.~\eqref{eq:G_I}.

\subsection{Effect of overlaps}
\label{sec:app_no_overlaps}
As shown in Eq.~\eqref{eq:finite_dim_overlap}, the autocorrelation function in general depends on two factors, the full eigenspectrum of the connectivity matrix and the overlaps between eigenvectors, both of which are modified when $\eta$ is changed. To disentangle the effects of the changes in the eigenspectrum and the changes in eigenvector overlaps, in this section we compare the autocorrelation we
derived in Sec.~\ref{sec:derivation_ac} with the autocorrelation we
would obtain if we assumed that the eigenvectors of $\vec{J}$ were
orthogonal. If that were the case, the autocorrelation is
computed as a sum of decoupled contributions associated with the
different eigenvalues and, in particular, the distribution of
eigenvectors would play no role in the result (see
Appendix~\ref{sec:app_no_overlaps} for
details). Figure~\ref{fig:four}a shows the predicted autocorrelations,
both including and excluding the contribution from the overlap \eqref{eq:overlap}. As expected, both predictions coincide for
$\eta=1$ and they increasingly depart from each other for decreasing
values of $\eta$. To better characterize this difference we show the
variance $C(0)$ as a function of the symmetry parameter for several
values of the spectral gap [Fig.~\ref{fig:four}b]. Remarkably, the
variance decreases with symmetry, but the opposite occurs when we
remove the contribution from eigenvector overlaps [see 'with' and
'without' overlap curves in Fig.~\ref{fig:four}b]. In both cases the
variance increases as spectral gaps get smaller, which is consistent
with the fact that the restoring drive towards the fixed point gets
weaker as the spectral gap gets smaller. This effect is, however, much
subtler when overlaps are not taken into
account~\citep{hennequin_vogels_gerstner_pre2012}.

The overlap also contributes to the overall time scale of the
autocorrelation, which we define by the quantity $\hat{\tau} = \int_0^{\infty}
t C(t)\,\d{t} / \int_{0}^{\infty} C(t)\,\d{t}$. This definition
guarantees that for an exponential autocorrelation $C(\tau) \propto
\exp(-|\tau|/\tau_0)$ the overall time scale is exactly $\tau_0$,
and provides a rough estimate of a natural time scale for autocorrelations
with more complex dependences. The numerical evaluation of $T$ shows
that the overall time scale is systematically smaller if the
contribution of the eigenvalues is removed
[Fig.~\ref{fig:four}c]. Unsurprisingly, either with or without the
overlap contribution the timescale gets longer as the spectral gap
gets smaller. Note also that the overall time scale $T$ varies
non-monotonically with the symmetry parameter [Fig.~\ref{fig:four}c],
unlike the asymptotic dependence shown in Fig.~\ref{fig:three}b.

\begin{figure}
 \centering
 \includegraphics{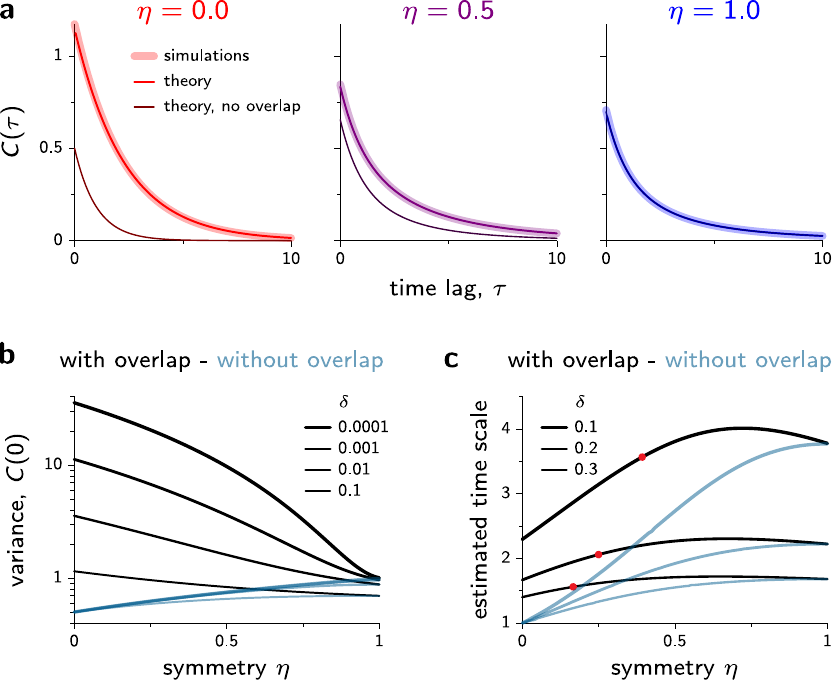}
 \caption{Effect of overlaps on autocorrelations and time scales. \textbf{a} Analytical predictions and numerical estimations of the average autocorrelation of network activity for $\delta=0.1$. We include two analytical predictions: the correct one, which takes properly into account the correlations among eigenvectors (`theory', based on Eq.~\eqref{eq:finite_dim_overlap} and its large-$N$ disorder-average limit, Eq.~\eqref{eq:Ctau_Adef}), and an incorrect prediction that assumes orthogonality among eigenvectors (`theory, no overlap'). \textbf{b} Theoretical prediction of the variance of network activity, given by $C(0)$, as a function of the symmetry parameter. As in \textbf{a}, we also include the prediction we would obtain if eigenvectors were orthogonal (`without overlap'). \textbf{c} Estimate of the overall time scale $\hat{\tau}$ (see text for details), with and without taking into account the effect of overlaps. Red dots indicate the transition between asymptotic regimes, as in Fig.~\ref{fig:three}b.}
 \label{fig:four}
\end{figure}

\section{Dynamics in the chaotic regime}

In the chaotic regime, the network generates its own fluctuating
activity without the need for external noise. Recall that chaotic
activity emerges as soon as the largest of the real parts of the
eigenspectrum, given by $-\delta$ and usually called spectral
abscissa, becomes positive. We follow the strategy of the preceding
section and we keep the spectral abscissa fixed while we vary the
symmetry parameter $\eta$.

\begin{figure*}
 \centering
 \includegraphics{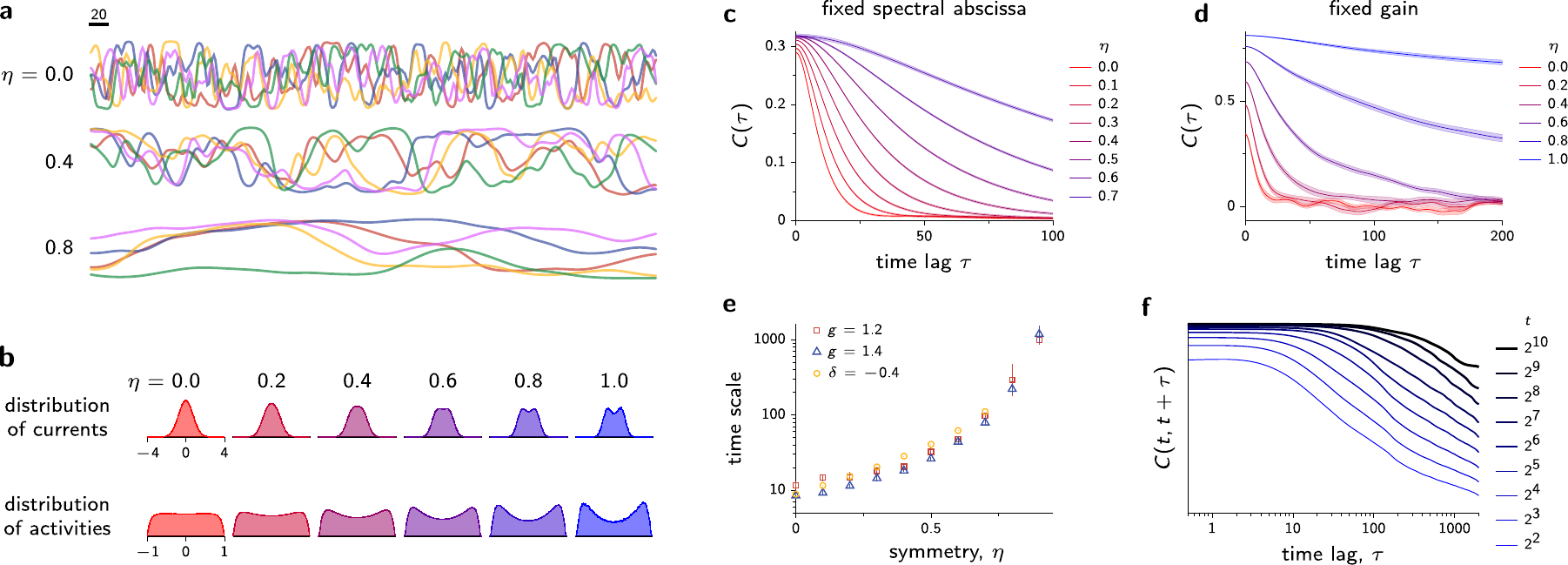}
 \caption{
 \label{fig:five}
   Effects of symmetry in the chaotic regime. \textbf{a} Firing rates of
 five arbitrary units in a network of size $N=10^4$, for three different values
 of $\eta$ (value indicated next to each inset). The spectral gap is $\delta=-0.4$ (i.e., the spectral abscissa is
 0.4). \textbf{b} Histograms of the
 currents $x$ and firing activities $\phi(x)$, for different values of $\eta$.
 Samples were taken every 40 time units from the simulated activity of a
 network of $10^{4}$ units for ten different realizations of the connectivity
 matrix, and $\delta=-0.4$.  \textbf{c} Population-average autocorrelation, for different values of
 the symmetry parameter $\eta$ and $\delta=-0.4$. The autocorrelation was
 estimated from the simulated activity of a network of $10^{4}$ units, using
 200 different draws of the connectivity matrix. The estimated standard error
 of the mean is shown in semitransparent shade. \textbf{d} Same as in \textbf{c},
 but keeping the gain fixed at $g=1.5$ instead of fixing $\delta$. For this
 panel we estimated the autocorrelation using ten independent realizations of
 the connectivity matrix. Notice the different axis ranges with respect to
 \textbf{c}. \textbf{e} Time scale of the network fluctuations, estimated from
 the width of the autocorrelation function at half of its maximum value, using
 the simulation results from \textbf{c} and \textbf{d}. The error bars indicate
 the standard error of the mean. \textbf{f} Non-stationarity of the network
 activity for $\eta=1$. The autocorrelation function depends on two time
 scales: the time lag $\tau$ and the time $t$ since the simulation started with
 an arbitrary initial condition. The autocorelation is estimated from simulated
 neuronal activity of size $N=10^4$ for $T=3030$ time units, using five different
 trials with different realizations of the connectivity matrix. In this figure the
 gain was fixed at $g = 1.4$.}
\end{figure*}

The evolution of firing activities shown in Fig.~\ref{fig:five}a suggests that in the chaotic regime the self-generated fluctuations get slower as $\eta$ increases. This slowing is accompanied by an increasing tendency of firing rates to linger around the extreme values of their dynamical range, as reflected by an increasingly bimodal distribution of currents $x$ and rates $\phi(x)$ when $\eta$ increases [Fig.~\ref{fig:five}b]. We quantified the slowing down of the fluctuations with the population-average autocorrelation. For $\eta=0$ the autocorrelation can be derived self-consistently in the limit of infinitely large networks, using the dynamical mean-field approach~(\citealt{sompolinsky_crisanti_sommers_prl1988,rajan_abbott_sompolinsky_pre2010}, see also Appendix~\ref{sec:app_DMF} for a general derivation). Unfortunately, this method does not lead to a closed-form solution for the autocorrelation as soon as $\eta >0$ (see Appendix~\ref{sec:app_DMF}) and we have to resort to numerical estimates, summarized in Fig.~\ref{fig:five}c for several values of $\eta$. For completeness we also include the autocorrelation functions for fixed gain, rather than fixed spectral abscissa [Fig.~\ref{fig:five}d].

The numerical estimates show that the time scale associated with the autocorrelation increases strongly as a function of $\eta$ and is considerably longer than in the fixed-point regime [Fig.~\ref{fig:five}e]. Such a slowing is rather insensitive to whether we fix the spectral abscissa or the gain, despite the fact that the variance $C(0)$ varies far more strongly when gain is fixed [Fig.~\ref{fig:five}d].

Quite strikingly, for $\eta=1$ fluctuations become slower as time goes by, and our initial assumption that the activity is stationary does not hold. The population-averaged autocorrelation $C(t, t+\tau)=\bigl[x(t) x(t + \tau)\bigr]_{J,N}$ at different points in time shows that the characteristic timescale of the autocorrelation grows with $t$ [Fig.~\ref{fig:five}d], a signature of aging dynamics \citep{bouchaud_etal_sprf1998}. For lower values of $\eta$, the dependence on the autocorrelation on the two timescales is less clear. Due to strong finite-size effects, it is difficult to determine from simulations alone whether aging appears also when the connectivity is not fully symmetric.

\section{Discussion}

In this work we examined the effect of partially symmetric connectivity on the
dynamics of randomly connected networks composed of rate units. We have derived
an analytical expression for the autocorrelation function in the regime of
linear fluctuations around the fixed point, and shown that increasing the
symmetry of the connectivity leads to a systematic slowing-down of the
dynamics. Numerical simulations confirm that a similar phenomenon takes place
in the chaotic regime of the network.

The impact of the degree of symmetry of the connectivity matrix on the
dynamics of neural networks has been a long-standing question in
theoretical neuroscience. Theorists initially focused on fully
symmetric networks of binary spin-like neurons \citep{hopfield_pnas1982} for
which tools from equilibrium statistical mechanics could be readily
applied \citep{amit_etal_pra1985}. After these initial studies, the realization
that brain networks are not symmetric led physicists to investigate
the dynamics of networks whose connectivity matrix has a random
antisymmetric component. It was found that departures from full
symmetry destroys spin-glass states, while retrieval states in
associative memory models were found to be robust to the presence of
weak asymmetry \citep{hertz_grinstein_solla_aip1986,crisanti_sompolinsky_pra1987,derrida_gardner_zippelius_europhyslett1987}.

Theorists also studied fully asymmetric networks, using rate models
\citep{sompolinsky_crisanti_sommers_prl1988}, networks of binary neurons
\citep{vanvreeswijk_sompolinsky_science1996} and networks of spiking neurons
\citep{brunel_jcns2000}. In all these models, chaotic states were shown to be
present for sufficiently strong coupling. In networks of spiking
neurons, chaotic states are characterized by strongly irregular
activity of the constituent neurons, with self-generated fluctuations
that evolve on fast time scales. Motivated by experimental findings,
recent studies have considered synaptic connectivity matrices where
bidirectionally connected pairs are overrepresented with respect to a
random network. In contrast with our model, in which no structure
exists beyond the level of pairs of neurons, these studies have
considered structured connectivity matrices in which partial symmetry
is a consequence of a larger-scale
structure. \citep{litwin-kumar_doiron_natns2012, stern_sompolinsky_abbott_pre2014} considered a
connectivity clustered into groups of highly connected neurons and
demonstrated that clustered connectivity could lead to slow
firing-rate dynamics generated by successive transitions between up
and down states within individual clusters.
An overrepresentation bidirectional connections can also arise in
networks with broad in- and out-degree distributions, which affect the
dynamics and the stability of asychronous states in such networks
\citep{roxin_ficn2011}. Other works have considered connectivities
with non-trivial second-order connectivity statistics, and studied the
resulting network dynamics. \citep{zhao_etal_ficns2011} analyzed how
the presence of connectivity patterns involving two connections (not
only bidirectionally connected pairs) affected the tendency for a
neuronal network to synchronize, while
\citep{bimbard_ledoux_ostojic_pre2016} focused on the oscillatory
activity generated by partially antisymmetric, delayed
interactions. Taking a completely different approach,
\citep{brunel_natns2016} showed that maximizing the number of patterns
stored in a network entails an overrepresentation of bidirectionnally
connected pairs of neurons, which suggests that partially symmetric
connectivity may be a signature of optimal information storage.

An important ingredient in our analysis is the fact that partially
symmetric interaction matrices are non-normal, i.e., they are not
diagonalizable by a set of mutually orthogonal eigenvectors. The
influence of non-normal connectivity on network dynamics has recently
received a considerable attention in the neuroscience
community~\citep{ganguli_sompolinsky,murphy_miller_neuron2009,goldman_neuron2008,hennequin_vogels_gerstner_pre2012,hennequin_vogels_gerstner_neuron2014,ahmadian_etal_pre2015}. Particularly
relevant to our study is the work by
\citep{hennequin_vogels_gerstner_pre2012}, who quantified the effects
of non-normality on the amplitude of the autocorrelation function in
random networks. Here we extend their results by studying the full
temporal shape of the autocorrelation function and by characterizing
how this shape is affected by the partial symmetry of connections.


The present work is also related to models of disordered systems and
spin glasses \citep{mezard_parisi_virasoro_spinglasses1987}. Most
studies in that field were inspired by physical phenomena and
considered fully symmetric interaction matrices. In that context, a
major result has been the discovery of aging, the phenomenon by
which dynamics become slower the longer the system
evolves~\citep{cugliandolo_kurchan_prl1993,cugliandolo_dean_pra1995,bouchaud_etal_sprf1998}. This
phenomenon has been observed in a broad class of complex systems
characterized by configuration spaces with extremely rugged energy
landscapes, composed of many local minima surrounded by high
barriers. In these systems a random initial condition is very likely
to set the system far from a stationary state and initiate a very slow
relaxation towards a fixed point. The relaxation takes infinitely long
for $N \rightarrow \infty$ because, loosely speaking, the longer the
system evolves, the deeper it wanders in the valleys of the energy
landscape, and the harder it becomes for it to find configurations of
lower energy \citep{bouchaud_etal_sprf1998}.

Whether fully symmetric interactions are necessary to observe aging
does not seem to be entirely understood, as to the best of our
knowledge only a few works seem to have considered partially symmetric
coupling~\citep{crisanti_sompolinsky_pra1987,iori_marinari_jpa1997,marinari_stariolo_jpa1998}.
Fully asymmetric networks have received more attention, but they do
not exhibit any aging phenomena. Here we interpolated between fully
asymmetric and fully symmetric networks, and have been able to obtain
mathematical results only in linear networks, in the non-chaotic
regime. Interestingly, we found that the partially symmetric case is
mathematically more complex than the symmetric or asymmetric
limits. This can be seen in the form of autocorrelation function
\eqref{eq:Adef_decomposition}, which simplifies considerably when
$\eta=0$ or $\eta=1$, but also in the Dynamical Mean Field Theory
(Appendix~\ref{sec:app_DMF}), where a coupling between the
autocorrelation and the response function appears for $\eta>0$. This
additional complexity results from the fact that the influence of a
single neuron's activity on all the other neurons is fed back through
couplings that are correlated with the neuron's activity, due to the
partial symmetry of the connections. More specifically, the inputs
received by neuron $i$ are given by terms $\sum_j J_{ij}\phi(x_j)$,
which are themselves influenced by the activity of neuron $i$. As a
result, neuron $i$ influences its own activity by an amount
proportional to the sum $\sum_j J_{ij}J_{ji} \phi(x_i)$, a random
number of mean $\eta \phi(x_i)$. The effect of this feedback loop is
that the individual input terms exhibit correlated fluctuations. When
$\eta=0$, the inputs received by neurons are uncorrelated and their
sum can be approximated by a Gaussian random variable whose mean and
variance can be determined
self-consistently~\citep{sompolinsky_crisanti_sommers_prl1988,rajan_abbott_sompolinsky_pre2010}. At
the other extreme, when $\eta=1$, the inputs received by neurons are
correlated, but the dynamics of the network can be described as a
relaxation of an energy function and the standard machinery of
statistical mechanics can be used. For other values of $\eta$, none of
these analytical strategies can be applied and the analysis becomes
more complex. Demonstrating analytically whether aging dynamics are
present in partially symmetric, non-linear networks seems an
outstanding open problem.

Our results on the autocorrelation function in the linear network are
closely related to recent results published by
\citep{bravi_sollich_opper_jpa2016}, who used a different set of
methods to compute the power spectrum of the network activity, i.e.,
the Fourier transform of the autocorrelation function of the same
model we investigated. Unlike~\citep{bravi_sollich_opper_jpa2016}, we
obtained the autocorrelation directly in real time, although our
results are fully consistent with theirs in that we obtain the same
two regimes with the same asymptotic timescales, depending on the
symmetry and the gain (or leak, in their case. cfr.\ Fig.~1 of
\citep{bravi_sollich_opper_jpa2016} with Fig.~\ref{fig:three}a).

Our work provides a potential bridge between two seemingly unrelated
observations in neuroscience. The first is the observation of strong
correlations between the synaptic strengths in pairs of cortical
pyramidal cells, the main excitatory neuronal type in cerebral cortex,
by multiple groups using in vitro electrophysiological recordings
\citep{markram_etal_jphysiol1997,song_etal_plosbiol2005,ko_etal_nature2011,perin_etal_pnas2011}. These
correlations are a consequence of two features of the connectivity:
First, there exists an overrepresentation of bidirectionally connected
pairs, compared to a Erd\H{o}s-R\'{e}nyi network with the same connection
probability.  For instance, \citet{song_etal_plosbiol2005} found a connection
probability of $c=0.116$ in pairs of neurons whose somas are less than
100 $\mu$m apart, while the probability that a pair of such neurons
are connected bidirectionally is approximately 4$c^2$. This degree of
overrepresentation has been found in multiple cortical areas, except
in barrel cortex where no such overrepresentation exists
\citep{lefort_etal_neuron2009}. Second, synaptic connections in bidirectionally
connected pairs are on average stronger than those in unidirectionally
connected pairs, and are significantly correlated
\citep{song_etal_plosbiol2005}. These observations lead to estimates
of $\eta \sim 0.5$, a value that, according to our model, would lead to a
significant increase in autocorrelation time scales compared to a
random asymmetric connectivity.

The second is the observation of long time scales in the
autocorrelations of neuronal activity from in vivo
electrophysiological recordings (see
e.g.~\citep{murray_etal_natns2014,huang_doiron_coinb2017}). Interestingly, the time scales of these
autocorrelations increase from sensory to higher level areas such as
the prefrontal cortex. Several mechanisms have been proposed to
account for this phenomenon: differences in the level of expression of
slow NMDA receptors \citep{wang_etal_pnas08}, or an increase in the strength of
recurrent connectivity \citep{elston_cercor2003} which could in particular lead to the presence of multiple fixed points that can slow down the dynamics \cite{litwin-kumar_doiron_natns2012, stern_sompolinsky_abbott_pre2014}. Our results suggest that this
increase in time scale could also be due to an increase in the degree
of symmetry of cortical connectivity. This would be consistent with
the study of \citep{wang_etal_natns2006}, who showed that the
overrepresentation of bidirectionally connected pairs of neurons is
significantly stronger in prefrontal cortex than in visual cortex.

From a neuroscience point of view, the model considered here is an
extremely simplified model of cortical networks because it lacks the
fundamental constraint that neurons are either excitatory or
inhibitory, and because it does not constrain firing rates to be
positive. These simplifications were made for the sake of mathematical
tractability. A few recent studies have investigated how these two
constraints influence the dynamics of such
networks~\citep{rajan_abbott_prl2006,ostojic_natns2014,kadmon_sompolinsky_prx2015,harish_hansel_ploscb2015,aljadeff_stern_sharpee_prl2015,mastrogiuseppe_ostojic_2016}. Extending
those works to connectivity with segregated excitation and inhibition
and partial symmetry is an important direction for future work that
might be facilitated by recent developments in random matrix theory
\citep{kuczala_sharpee_pre2016,aljadeff_renfrew_stern_jmp2015}.

\subsubsection*{Acknowledgments}

We thank Johnatan Aljadeff for his comments on a previous version of the manuscript.
The research leading to these results has received funding from the People Programme (Marie Curie
Actions) of the European Union's Seventh Framework Programme FP7/2007--2013/ under REA grant
agreement 301671. This has also been funded by the Programme Emergences of City of Paris, and the program
``Investissements d'Avenir'' launched by the French Government and
implemented by the ANR, with the references ANR-10-LABX-0087 IEC and
ANR-11-IDEX-0001-02 PSL* Research University.

The funders had no role in study design, data collection and analysis, decision to publish, or preparation of the manuscript.

\appendix
\small
\section{Derivation of the double complex integral}
\label{sec:app_CM}
We summarize here the derivation of the double complex integral of Eq.~\eqref{eq:Ctau_Adef}. Before doing so, we sketch the derivation of the local density of the overlap done in~\citep{chalker_mehlig_prl1998}, as this will let us introduce some notation and pave the way for our calculation.

For a complex variable $z=x + \I y$, with $x$ and $y$ real and with
conjugate $\bar{z}=x - \I y$, we define the Wirtinger derivatives
$\partial/\partial z = (\partial / \partial x - \I \partial/
\partial y) / 2$ and $\partial/\partial \bar{z} = (\partial / \partial x +
\I \partial/ \partial y) / 2$, which obey $\partial z / \partial z =
\partial \bar{z} / \partial \bar{z} = 1$ as well as $\partial \bar{z}/
\partial{z} = \partial z / \partial \bar{z} = 0$. The complex
differential is defined to be $\dd{z} \equiv \d{z}\,\d{\bar{z}} = 2\,\d{x}\,\d{y}$,
where the factor 2 comes from the Jacobian. We also define the complex Dirac delta
so that it obeys the relation $\int \delta^{2}(z)\,\dd{z} \equiv 1$, which
implies $\delta^{2}(z) = (1/2) \delta(x)\delta(y)$ given our
convention for the complex differential. Two useful identities for the $\delta$ function in the complex plane are
\begin{equation}
 \delta^{2}(z) = \frac{1}{2\pi} \frac{\partial}{\partial \bar{z}}
 \frac{1}{z} = \frac{1}{2\pi} \frac{\partial}{\partial z} \frac{1}{\bar{z}},
 \label{eq:identity_deltaz}
\end{equation}
which can be checked by integrating over $\dd{z}$ and applying the
complex version of Green's theorem:
\begin{equation}
\int \left(\frac{\partial v}{\partial z} + \frac{\partial \bar{v}}{\partial \bar{z}}\right)\,\dd{z} = \I \oint (v\,\mathrm{d}\bar{z} - \bar{v}\,\mathrm{d}z),
\label{eq:green_th}
\end{equation}
where $v$ and $\bar{v}$ are to be considered independent functions.

The resolvent, defined for any matrix $\mathbf{J}$ as $(z\mathbf{1} - \mathbf{J})^{-1}$, is a key quantity in the analysis of random matrices because it can be ensemble-averaged using standard methods and can be related to quantities of interest. So, for example, the empirical density of eigenvalues of a given $\mathbf{J}$,
\[ \rho_J(z) = \frac{1}{N} \sum_{i=1}^{N} \delta(x - \Re \lambda_i)
\delta(y - \Im \lambda_i) = \frac{2}{N} \sum_{i=1}^{N} \delta^2(z -
\lambda_i),\]
can be expressed thanks to the identities~\eqref{eq:identity_deltaz} as
\[ \rho_J(z) = \frac{1}{\pi} \frac{\partial}{\partial \bar{z}}
 \frac{1}{N} \sum_{i=1}^{N} \frac{1}{z - \lambda_i} = \frac{1}{\pi} \frac{\partial}{\partial \bar{z}}
\frac{1}{N} \Tr (z\mathbf{1} - \mathbf{J})^{-1}.\]
This quantity is hard to compute for any particular realization at
finite $N$, but it becomes easier to handle in the limit of large $N$,
where all empirical densities converge to the average density
\[ \rho(z) = \left[ \rho_{J}(z) \right]_J = \frac{1}{\pi} \frac{\partial}{\partial \bar{z}}
\left[ \frac{1}{N} \Tr (z\mathbf{1} - \mathbf{J})^{-1} \right]_{J}.\]
Deriving the average density, therefore, amounts to computing the function $G(z) = \left[\Tr (z\mathbf{1} -
\mathbf{J})^{-1} / N\right]_{J}$ in the large $N$ limit.

The local
density of the overlap can be derived in a similar manner using the
spectral decomposition $(z\mathbf{1} - \mathbf{J})^{-1} = \sum_{i=1}^{N}
\mathbf{R}_{i} (z - \lambda_i)^{-1} \mathbf{L}^{\dag}_{i}$, where
$\vec{R}_i$ and $\vec{L}_i$ are the right and left eigenvectors
of $\vec{J}$, respectivelyt. If we substitute the definition of the overlap
matrix into Eq.~\eqref{eq:densityD} and we use the
identities~\eqref{eq:identity_deltaz} we obtain
 \begin{align}
 D(z_1, z_2) & = \frac{4}{N} \left[ \sum_{i,j=1}^{N}(\vec{L}^{\dag}\vec{L})_{ij}
(\vec{R}^{\dag}\vec{R})_{ji} \delta^{2}(z_1 -
\lambda_i) \delta^{2}(z_2 - \lambda_j) \right]_{\!J}\notag \\
& = \frac{1}{N\pi^2} \frac{\partial}{\partial \bar{z}_1}
\frac{\partial}{\partial z_2} \left[ \sum_{i,j=1}^{N} \Tr\, \vec{R}_i
 \frac{1}{z_1 - \lambda_i}\vec{L}^{\dag}_{i}
 \vec{L}_{j} \frac{1}{\bar{z}_2 - \bar{\lambda}_j} \vec{R}^{\dag}_j
 \right]_{\!J} \notag \\
& = \frac{1}{\pi^2} \frac{\partial}{\partial \bar{z}_1}
\frac{\partial}{\partial z_2} \left[ \frac{1}{N} \Tr\,
 \frac{1}{z_1\mathbf{1} - \mathbf{J}}
 \frac{1}{\bar{z}_2\mathbf{1} - \mathbf{J}^{\dag}}\right]_{\!J},
 \label{eq:DddG}
\end{align}
and the problem reduces to computing the quantity
\[G(z_1, z_2) = \left[ \frac{1}{N} \Tr\,
 \frac{1}{z_1\mathbf{1} - \mathbf{J}}
 \frac{1}{\bar{z}_2\mathbf{1} - \mathbf{J}^{\dag}}\right]_{\!J}.
\]
The expression of $G(z_1, z_2)$ for the ensemble of Gaussian random matrices with partial symmetry was derived by
~\citet{mehlig_chalker_jmathphys2000}. The basic idea behind their calculation is to expand resolvents in power series, average over the disorder term by term, and organize the sums so that a recursive relation can be established and ultimately solved (for a thorough description of the method, see also~\citep{ahmadian_etal_pre2015}).
The result is a complex function that takes the value
\begin{multline}
 G(z_1, z_2) =
\frac{1}{1 - \eta^2} \\
\times \left( \frac{(1 - \eta^2)^{2} + \eta (z_1^2 + \bar{z}_2^2) - (1 + \eta^2) z_1 \bar{z}_2}{|z_{1} - z_{2}|^{2}} - 1\right)
\label{eq:GinsideE}
\end{multline}
when both $z_1$ and $z_2$ lie inside the ellipse centered at the origin and which has major and minor radii $1 + \eta$ and $1 - \eta$, respectively. We will call this ellipse $E_\eta$ for later convenience. When $z_1$ and $z_2$ lie outside $E_\eta$ we have instead
\begin{equation*}
G(z_1, z_2) = \dfrac{h_1 \bar{h}_2}{1 - h_1 \bar{h}_2},
\end{equation*}
where
\begin{equation*}
 h_1 = \frac{z_1 - \sqrt{z_1^2 - 4\eta}}{2\eta},\qquad
 \bar{h}_2=\frac{\bar{z}_2 - \sqrt{\bar{z}_{2}^{2} - 4\eta}}{2\eta}.
\end{equation*}
Right on the ellipse $E_\eta$, $|h_i| = 1$. When both $z_1$ and $\bar{z}_2$ lie outside the ellipse, the function $G(z_1, z_2)$ is analytic on $z_1$ and $\bar{z}_2$. This analyticity implies, from \eqref{eq:DddG}, that the local density of the overlap vanishes outside the ellipse.

We now proceed to compute $A(u, \tau)$ [Eq.~\eqref{eq:Adef}]. Inserting the identity~\eqref{eq:DddG} into \eqref{eq:Adef} leads to
\[ A(u, \tau) = \frac{1}{4\pi^2} \iint \E^{g(z_1 + \bar{z}_2) u + g z_1 \tau} \frac{\partial}{\partial \bar{z}_1} \frac{\partial}{\partial z_2} G(z_1, z_2)\,\dd{z_1} \dd{z_2}.\]
Because the exponential prefactor is analytic in $z_1$ and $\bar{z}_2$,
it commutes with the two partial derivatives. We can therefore
apply Green's theorem twice to obtain
\begin{equation}
 A(u, \tau) = \frac{1}{4\pi^2} \oint_{\!E} \oint_{\!E} \E^{g (z_1 + \bar{z}_2) u + g z_1
\tau} G(z_1, z_2)\,\d{z_1}\d{\bar{z}_2},
\label{eq:Iutau}
\end{equation}
where both contour integrals are around the ellipse $E$, at whose boundary $G(z_1, z_2)$ stops being analytic. To compute $A(u, \tau)$ we follow the approach of \citep{mehlig_chalker_jmathphys2000} and use the linear transformation $w=(z - \eta \bar{z}) / (1 - \eta^2)$ (or, equivalently, $z = w + \eta \bar{w}$) to reshape the contour of integration from the ellipse $E_{\eta}$ into the unit circle. Applying this transformation to both $z_1$ and $z_2$, the surface integrals in~\eqref{eq:Iutau} become countour integrals on the unit circle $|w|^2=w \bar{w} = 1$. On this contour we can replace every $\bar{w}$ in the integrand by $w^{-1}$ and we can use the standard tools of complex analysis to carry out the integrals. We describe in more detail our derivation in the following.

\begin{figure}
 \centering
 \includegraphics{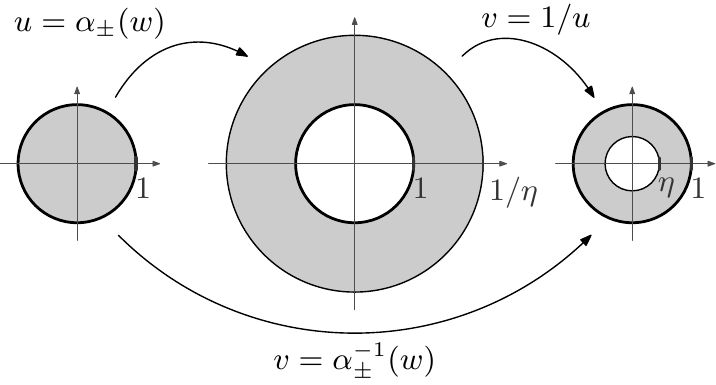}
 \caption{
 \label{fig:domain_alpha}
   Mapping of the unit disk $D = \{w \in \mathbf{C} \mid\; |w| \leq 1\}$ by the functions $\alpha_{\pm}(w)$ and $\alpha^{-1}_{\pm}(w)$. }
\end{figure}

We start by performing the integral over $\bar{z}_2$, expressing Eq.~\eqref{eq:GinsideE} in terms of $w_1$, $\bar{w}_1$, and $w_2$, and replacing all $\bar{w}_2$ by $w_2^{-1}$. After some simplifications, we obtain
\begin{multline}
 G\bigl(z_1(w_1), z_2(w_2)\bigr) = -1 + \frac{1}{\eta(\alpha_{+} - \alpha_{-})}\\
 \times \left( \frac{1 - \eta w_1 \alpha_{+}}{w_2 - \alpha_{+}} - \frac{1 - \eta w_1 \alpha_{-}}{w_2 - \alpha_{-}}\right),
\label{eq:Gtilde_eta}
\end{multline}
where we defined the poles
\begin{equation}
 \alpha_{\pm}(w_1) = \frac{\bar{w}_1 + \eta w_1 \pm \sqrt{(\bar{w}_1 + \eta w_1)^2 - 4\eta}}{2 \eta}.
 \label{eq:roots}
\end{equation}
These poles depend on $w_1$ and and can be shown to map the unit disk onto an annulus of inner radius 1 and outer radius $1/|\eta|$ [Fig.~\ref{fig:domain_alpha}].
This information will be relevant when we use residue calculus.

The double integral \eqref{eq:Iutau} has to be regularized because the integrand diverges at $z_1 = z_2$. Our regularization consists of first integrating $w_2$ on the unit circle while constraining $w_1$ to be on a concentric circle of smaller radius $|w_1| = 1 - \epsilon$, with $\epsilon>0$ small. Once the integral over $w_2$ is done, we take the limit $\epsilon \rightarrow 0$ and perform the second integral over $w_1$.

Under this regularization, we decompose the double integral~\eqref{eq:Iutau} as
\begin{multline}
 A(u, \tau) = \lim_{\epsilon \rightarrow 0} \frac{1}{4\pi^2} \oint_{|w_1|=1 - \epsilon} \!\!\E^{g (w_1 + \eta \bar{w}_1)(u + \tau)} \mathcal{A}(w_1, u)\\ \times (\d{w}_1 + \eta \d{\bar{w}}_1)
\label{eq:Autau}
\end{multline}
where we used $z_1 = w_1 + \eta \bar{w}_1$ and we defined
\begin{multline} \mathcal{A}(w_1, u) = \oint_{|w_2|=1} \!\!\E^{g (w^{-1}_2 + \eta w_2)u}
 G\bigl(z_1(w_1), z_2(w_2)\bigr)\\
 \times \,\left(-\frac{1}{w_2^2} + \eta\right) \d{w_2}.
 \label{eq:inner_w2}
\end{multline}
Note that here we used $\bar{w}_2 = w_2^{-1}$ to express the integrand and the differential $\d{\bar{z}}_2 = \d{\bar{w}}_2 + \eta \d{w}_2$ in terms of $w_2$ only. The integrand of Eq.~\eqref{eq:inner_w2} contains one singularity inside the contour of integration, at $w_2=0$. This singularity is associated with the essential singularity from the exponential, $\E^{1/w_2}$, as well as with the pole of second order $1 / w_2^{2}$. Because we are assuming that $|w_1| < 1$, the poles of $G\bigl(z_1(w_1), z_2(w_2)\bigr)$ at $w_2=\alpha_{\pm}$ lie outside the contour and therefore do not contribute to the integral. We are thus left with the task of computing the residue at the origin. We do that by expanding the integrand in Laurent series around $w_2 = 0$, using the relations
 \begin{gather*}
 \E^{(z^{-1} + \eta z)t} = \sum_{k=-\infty}^{\infty} \left( z \sqrt{\eta} \right)^{k} I_{k}(2\sqrt{\eta}t),\quad \text{for } z \neq 0,\\
 (z - z_0)^{-1} = -\frac{1}{z_0} \sum_{k=0}^{\infty} \left(\frac{z}{z_0} \right)^{k},\quad \text{for } |z| < |z_0|,
 \end{gather*}
 with $I_k(z)$ being the modified Bessel function of order $k$. We use the last power series to expand the terms $(w_2 - \alpha_{\pm})^{-1}$ in $G\bigl(z_1(w_1), z_2(w_2)\bigr)$ [Eq.~\eqref{eq:Gtilde_eta}]. This power series converges because $|w_2| < |\alpha_{\pm}|$ when $|w_1|<1$, as we assume in our regularization scheme. After expanding, applying Cauchy's residue theorem, and taking the limit $\epsilon \rightarrow 0$, we obtain
 \begin{equation*}
 \mathcal{A}(w_1, u) = -\frac{2\pi\I}{gu} \sum_{k=0}^{\infty} w_{1}^{-k} \eta^{-k/2}k\, I_{k} (2g\sqrt{\eta}u),
 \end{equation*}
The final step is to compute the integral in \eqref{eq:Autau} with the same strategy we used for $\mathcal{A}(w_1, u, \tau)$. In this case we express the integrand in terms of $w_1$ only and we expand the exponential factor in \eqref{eq:Autau} with the identity
\[ \E^{(z+ \eta z^{-1})t} = \sum_{k=-\infty}^{\infty} \left( \frac{z}{\sqrt{\eta}} \right)^{k} I_{k}(2\sqrt{\eta}t).\]
We then pick the residue from the expansion and apply Cauchy's theorem. The result is
\begin{equation*}
 A(u, \tau) = A_1(u, \tau) + A_2(u, \tau),
\end{equation*}
with
\begin{multline}
A_1(u, \tau) =
\sum_{k=-\infty}^{\infty} \eta^{k/2} I_k(2g\sqrt{\eta}\tau) \biggl[ (1 + \eta^2) I_k\bigl(2g (1 + \eta) u\bigr)\\ - \eta
\bigl(I_{k-2}\bigl(2g (1 + \eta) u\bigr) + I_{k+2}\bigl(2 g(1 +
\eta) u\bigr)\bigr) \biggr]
\label{eq:A1pre}
\end{multline}
and $A_2(u, \tau)$ given by Eq.~\eqref{eq:A2}.

The expression~\eqref{eq:A1pre} for $A_1(u, \tau)$ can be further simplified with the identity \citep{prudnikov_etal}
\begin{multline}
 \sum_{k=-\infty}^{\infty} \E^{\I k\alpha} J_{k}(w) J_{k+\nu} (z) = \left( \frac{z - w\E^{-\I\alpha}}{z - w\E^{\I\alpha}}
 \right)^{\nu/2} \\ \times J_{\nu}\left(\sqrt{w^2 + z^2 - 2wz\cos \alpha}\right),
 \label{eq:prudnikov_orig}
\end{multline}
which we can transform into a more convenient expression for our problem, using $J_\nu(\I z) = \I^{\nu} I_{\nu}(z)$ and taking $\alpha = \pi - (\I / 2) \ln \eta$ so that
$\E^{\I \alpha} = -\sqrt{\eta}$. The identity~\eqref{eq:prudnikov_orig} then becomes
\begin{multline}
 \sum_{k=-\infty}^{\infty} \eta^{k/2} I_{k}(w) I_{k+\nu} (z) = \left( \frac{z + w\eta^{-1/2}}{z + w \eta^{1/2}}
 \right)^{\nu/2}\\ \times I_{\nu}\left(\sqrt{w^2 + z^2 + wz(\eta^{1/2} + \eta^{-1/2})}\right),
 \label{eq:prudnikov}
\end{multline}
which allows us to rewrite Eq.~\eqref{eq:A1pre} as the final expression~\eqref{eq:A1}.

The series $A_2(u, \tau)$ does not seem to have a closed expression for general $\eta$. For $\eta=1$, however, we can
exploit the identity
\[ \sum_{k=-\infty}^{\infty} I_k(w) I_{n - k}(z) = I_n(w + z)\]
to conclude that
\[A_{2}(u, \tau) = -I_0\bigl(2g(2u + \tau)\bigr) + I_2\bigl(2g(2u + \tau)\bigr) \quad\text{(for $\eta=1$)}.\]

\section{Evaluation of the time integral for long $\tau$}
\label{sec:app_laplace}
\begin{figure}
  \includegraphics{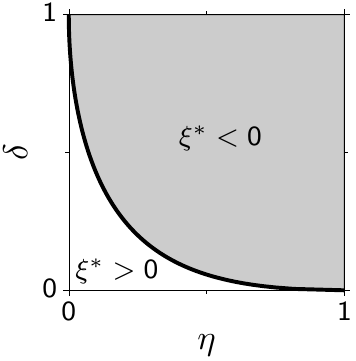}
  \caption{
  \label{fig:flip_phi}
    Sign of $\xi^{*}$, at which the exponent in Eq.~\eqref{eq:finalCtau} is largest, as a function of the spectral gap and the degree of symmetry. This diagram is equivalent to that shown in Fig.~\ref{fig:three}a, after transforming $\delta$ into its associated gain $g = (1 - \delta)/(1 + \eta)$.}
\end{figure}

The exact average autocorrelation is given in~\eqref{eq:Ctau_Adef} as a time integral that can be decomposed as
\[C(\tau) = \int_{0}^{\infty} \E^{-2u - \tau} \bigl[A_1(u, \tau) + A_2(u, \tau)\bigr]\, \d{u},\]
where $A_1(u, \tau)$ and $A_2(u, \tau)$ are defined in Eqs.~\eqref{eq:A1} and \eqref{eq:A2}. To gain more analytical insight we will evaluate $C(\tau)$ when $\tau$ is sufficiently large, a limit that allows us to invoke Laplace's method and approximate the integral with a closed-form expression~\citep{bender_orszag_1999}. Before applying the limit it is convenient to express this integral in terms of a new variable $\xi\equiv u / \tau$, which is well defined for $\tau > 0$. With this definition
\begin{equation}
 \begin{split}
 C(\tau) &= \int_{0}^{\infty} \E^{-\tau (2\xi + 1)} \bigl[A_1\bigl(u(\xi), \tau\bigr) + A_2\bigl(u(\xi), \tau\bigr)\bigr]\, \tau\d{\xi},\\
 & \equiv C_1(\tau) + C_2(\tau).
 \end{split}
 \label{eq:AC_xi}
\end{equation}
which we split into the two terms composing the integrand. We start with the asymptotic dependence of $C_1(\tau)$, ignoring for the moment $C_2(\tau)$. The integrand of $C_1(\tau)$ contains $I_0(\cdot)$ and $I_2(\cdot)$, whose argument is large in the long-$\tau$ limit. We can therefore use the asymptotic expansion of the modified Bessel functions of order $\nu$,
\begin{equation}
 I_{\nu}(x) = \frac{\E^{x}}{\sqrt{2\pi x}} \left[1 - \frac{4\nu^2 - 1}{8x} + O(x^{-2})\right]\quad\text{for}\ x \gg 1.
 \label{eq:Besselexp}
\end{equation}
At this order, defining
\[\psi(\xi;\eta) \equiv \psi\bigl(u(\xi), \tau; \eta\bigr) / \tau = 2 \sqrt{(1 + \eta)^2 \xi (\xi + 1) + \eta},\]
we obtain
\begin{multline}
 C_1(\tau) = \sqrt{\frac{\tau}{2\pi}} \int_{0}^{\infty} \frac{\exp\bigl\{ -\tau \left[ (2\xi + 1) - g \psi(\xi;\eta) \bigr]\right\}}{\sqrt{g \psi(\xi; \eta)}} \\
 \Biggl[ (1 + \eta^2)\left(1 + \frac{1}{8 \tau g \psi(\xi; \eta)} \right)\\ -2 \eta \left(1 + \frac{2(1 -\eta)^2}{\psi(\xi; \eta)^2} \right)
\left(1 - \frac{15}{8 \tau g \psi(\xi; \eta)} \right) \biggr] \, \d{\xi}.
\label{eq:finalCtau}
\end{multline}
with $g = (1 - \delta) / (1 + \eta)$.
This integral is of the form
\begin{equation}
A(\tau) = \int_{0}^{\infty} f(\xi) \E^{\tau b(\xi)}\,\d{\xi}.
\label{eq:laplaceI}
\end{equation}
In the limit of large $\tau$, only the infinitessimal interval around the maximum of $b(\xi)$ contributes to the integral because the contribution of the remaining intervals is exponentially suppressed. In our particular case the maximum of $b(\xi)$ is at
\begin{equation}
 \xi^{*} = \frac{1}{2} \left( -1 + \frac{1 - \eta}{(1 + \eta)\sqrt{2\delta - \delta^2}}\right).
 \label{eq:xistar_tilde}
\end{equation}
We distinguish two cases. For large enough values of $\eta$ and $\delta$, $\xi^{*}$ is negative [Fig.~\ref{fig:flip_phi}], which means that in the integration range $[0, \infty)$ of Eq.~\eqref{eq:finalCtau} the maximum value of $b(\xi)$ is at $\xi=0$. Conversely, for low values of $\eta$ and $\delta$, the maximum of $b(\xi)$ occurs within $(0, \infty)$. These two cases will lead to different time dependences and will be studied separately in the following.

For values of $\eta$ and $\delta$ such that $\xi^{*} < 0$, the limit $\tau \rightarrow \infty$ of Eq.~\eqref{eq:laplaceI} can be approximated by~\citep[pp.~266-268]{bender_orszag_1999}
\begin{align}
 A(\tau) & \sim \lim_{\epsilon \rightarrow 0} \int_{0}^{\epsilon} f(0)\E^{\tau[b(0) + b'(0) s]}\,\d{s}\notag \\
 & \sim \int_{0}^{\infty} f(0) \E^{\tau[b(0) + b'(0) s]}\,\d{s}
 \sim -\frac{f(0)\E^{\tau b(0)}}{\tau b'(0)},
 \label{eq:laplaceI_bdr}
\end{align}
where $b'(0)$ denotes the derivative of $b(\xi)$ evaluated at $\xi^{*}=0$.
Applying this approximation to \eqref{eq:finalCtau}, we obtain
\begin{multline}
 C_1(\tau) = \frac{\tau^{-3/2} \eta^{-1/4}}{2\sqrt{\pi}} \frac{1 + \eta^2}{\delta(1 + \eta) - [1 - \sqrt{\eta}]^2} \left(\frac{1 + \eta}{1 - \delta}\right)^{3/2}\\
 \times \exp\left\{-\tau \frac{(1 - \sqrt{\eta})^2 + 2\delta \sqrt{\eta}}{1 + \eta} \right\}
 \label{eq:C1_xi0}
\end{multline}
Conversely, if $\eta$ and $\delta$ are such that $\xi^{*} > 0$, the maximum $\xi^{*}$ falls within the integration region and we can approximate the large-$\tau$ limit of the integral~\eqref{eq:laplaceI} by~\citep[see][p.~267]{bender_orszag_1999}
\begin{multline*}
A(\tau) \sim \lim_{\epsilon \rightarrow 0} \int_{\xi^{*} - \epsilon}^{\xi^{*} + \epsilon} f(\xi^{*})\E^{\tau(b(\xi^{*}) + (s - \xi^{*})^2 b''(\xi^{*}) / 2)}\,\d{s}\\
\sim \int_{-\infty}^{\infty} f(\xi^{*})\E^{\tau(b(\xi^{*}) + (s - \xi^{*})^2 b''(\xi^{*}) / 2)}\,\d{s}\\
\sim \frac{\sqrt{2\pi} f(\xi^{*})\E^{\tau b(\xi^{*})}}{\sqrt{-\tau b''(\xi^{*})}}.
\end{multline*}
with $b''(\xi^{*})$ denoting the second derivative of $b(\xi)$ evaluated at $\xi^{*}$.
In this case Eq.~\eqref{eq:finalCtau} is approximately given by
\begin{equation*}
 C_1(\tau) = \frac{\delta^{-1/2}(1 - \eta)^2}{2\sqrt{2} (1 - \delta)} \exp\left\{-\tau \frac{1 - \eta}{1 + \eta} \sqrt{2\delta - \delta^2}\right\}.
\end{equation*}

We now turn to the asymptotic dependence of $C_2(\tau)$.
The original form for $C_2(\tau)$ is
\begin{multline}
 C_2(\tau) = \int_{0}^{\infty} \biggl\{ \E^{-\tau (2\xi + 1)} \sum_{k=0}^{\infty} \eta^{k+2} (k+1)^2
 \\ \times \frac{I_{k + 1} \bigl(2 g \tau \sqrt{\eta} \xi\bigr)
I_{k+1} \bigl(2 g \tau \sqrt{\eta}(\xi + 1)\bigr)}{\eta \tau g^2 \xi (\xi + 1)}\biggr\} \d{\xi}.
\label{eq:danglingterm}
\end{multline}
The integrand contains a product of Bessel functions that grows exponentially with $\tau$ (see Eq.~\eqref{eq:Besselexp}), but this growth is kept in check by the exponential prefactor. To see this, we introduce
 the scaled modified Bessel function,
\[ I_{\nu}^{*}(x) \equiv \E^{-x} I_{\nu}(x),\]
in terms of which~\eqref{eq:danglingterm} becomes
 \begin{multline}
 C_2(\tau) = \int_{0}^{\infty}
 \exp\left[-\tau \frac{(1 - \sqrt{\eta})^2 + 2\delta \sqrt{\eta}}{1 + \eta} \right]
 \sum_{k=0}^{\infty} \eta^{k+1} (k+1)^2 \\
 \times \frac{I^{*}_{k + 1} \bigl(2 g\tau \sqrt{\eta}\xi \bigr)
 I^{*}_{k+1} \bigl(2 g \tau \sqrt{\eta}(\xi + 1)\bigr)}{\tau g^2 \xi(\xi+1)}\,\d{\xi}.
\label{eq:danglingterm_xi}
 \end{multline}
 The integral converges because the exponential decays to zero for large $\tau$ and trumps the power-law decay of $I^{*}_k(\cdot)$. Equation~\eqref{eq:danglingterm_xi}
 also has the same form as \eqref{eq:laplaceI} and contains exactly the same exponent as in Eq.~\eqref{eq:C1_xi0}; because the exponential attains its maximum at $\xi=0$, we can use the approximation~\eqref{eq:laplaceI_bdr}. To evaluate at $\xi=0$ we use the power expansion $I^{*}_{k + 1} \bigl(2 g \tau \sqrt{\eta}\xi\bigr) \stackrel{\xi \sim 0}{=} (g \tau \sqrt{\eta} \xi)^{k+1}$, and identify $f(\xi)$ in \eqref{eq:laplaceI_bdr} with
\begin{equation*}
 f(\xi) \equiv \sum_{k=0}^{\infty} \frac{\eta^{3(k + 1)/2}}{\xi + 1} (k+1)^2 g^{k-1} \xi^{k} \tau^{k}
 I^{*}_{k+1}\bigl(2g \tau \sqrt{\eta} (\xi + 1)\bigr).
\end{equation*}
All the terms of $f(\xi)$ with $k>0$ vanish at $\xi=0$, and Eq.~\eqref{eq:laplaceI_bdr} reads in this case
\begin{multline*}
 C_2(\tau)
 \sim \frac{\tau^{-3/2}\eta^{5/6}}{4\sqrt{\pi}} \frac{(1 + \eta)^{5/2}}{(1 - \delta)^{3/2}} \frac{1}{(1 - \sqrt{\eta})^2 + 2 \sqrt{\eta}\delta}\\
 \times \exp\left\{ -\tau \frac{(1 - \sqrt{\eta})^2 + 2\delta \sqrt{\eta}}{1 + \eta} \right\}
\end{multline*}
Summing this contribution to that in Eq.~\eqref{eq:C1_xi0} leads to the result reported in Eqs.~\eqref{eq:C_asymp_powerlaw}--\eqref{eq:G_II}.

\section{Autocorrelation without overlaps}
\label{sec:app_nooverlap}
Here we compute the autocorrelation ignoring the effect of the overlaps between eigenvectors. In this case, we need to compute the individual contribution of a single eigenvalue to the autocorrelation, and then sum over the contributions of all eigenvalues. We start with the one-dimensional version of Eq.~\eqref{eq:linear}
\begin{equation}
 \frac{\d{x}}{\d{t}} = \alpha x + \sigma \xi(t),
 \label{eq:onedim}
\end{equation}
where the parameter $\alpha$ would be the (single) eigenvalue of the system, assumed to have negative real part to prevent $x(t)$ to grow unbounded, and where $\xi(t)$ is a source of standard Gaussian white noise. The solution of \eqref{eq:onedim} is
\[ x(t) = \sigma \int_{-\infty}^{t} \E^{\alpha (t - s)}\xi(s)\,\d{s}, \]
from which we can derive the autocorrelation:
\begin{align}
 \avg{x(t) x(t + \tau)} & = \sigma^2 \int_{-\infty}^{t} \int_{-\infty}^{t + \tau} \E^{\alpha (2t + \tau - s - u)}\avg{\xi(s)\xi(u)} \,\d{s}\,\d{u}\notag \\
 & = - \sigma^{2} \frac{\E^{\alpha \tau}}{2\alpha} \equiv \sigma^{2} C_\alpha(\tau). \label{eq:single_C}
\end{align}
The eigenvalue $\alpha$ determines both the time scale and the amplitude of the autocorrelation. We set the overall factor $\sigma^2$ to 1, without loss of generality.

The average autocorrelation for the high-dimensional system in the absence of overlaps is the sum of \eqref{eq:single_C} over all the eigenvalues. In the large-$N$ limit we would have
\begin{equation}
 C(\tau) = \frac{1}{N} \sum_{i=1}^{N} C_{\alpha_i}(\tau) \xrightarrow[]{N\rightarrow \infty} \int C_{\alpha}(\tau)\rho(\alpha)\,\d{\alpha}\d{\bar{\alpha}},
 \label{eq:C_as_avg_alpha}
\end{equation}
where $\rho(\alpha)$ is the probability density of eigenvalues and the integral is on the complex plane. For the system~\eqref{eq:linear} and for the connectivity matrices we consider, the density of the eigenvalues $\alpha$ is uniform and has support on an ellipse centered at $z=-1$ with major radius $g(1 + \eta)$ and minor radius $g(1 - \eta)$. The integral~\eqref{eq:C_as_avg_alpha} can be computed in that case and reads
\begin{equation}
 C(\tau) = \frac{-1}{\pi g^2 (1 - \eta^2)} \int_{E} \frac{\E^{\alpha \tau}}{2\alpha}\,\d{\alpha}\d{\bar{\alpha}},
 \label{eq:Ctau_indep}
\end{equation}
where we used Eq.~\eqref{eq:single_C} and the prefactor is the constant value that $\rho(\alpha)$ takes on the elliptic support $E$. To evaluate the integral we use the parametrization
\[\alpha = -1 + r (1 + \eta) \cos \theta + \I r (1 - \eta) \sin \theta\]
and integrate over $r \in [0, g]$ and $\theta \in [0, 2\pi]$. Noting that
\[ \int_{E} \d{\alpha}\d{\bar{\alpha}} = (1 - \eta^2) \int_{0}^{2\pi} \int_{0}^{g} r\, \d{r}\,\d{\theta},\]
Eq.~\eqref{eq:Ctau_indep} becomes
\begin{equation}
 C(\tau) = \frac{1}{\pi g^2} \int_{0}^{2\pi}\int_{0}^{g} \frac{\exp\bigl\{-\tau\bigl[1 - r \psi (\theta) \bigr]\bigr\}}{2 [1 - r \psi(\theta)]} r\, \d{r}\,\d{\theta}.
 \label{eq:Ctaucplx}
\end{equation}
where for convenience we defined
\[\psi(\theta) \equiv (1 + \eta)\cos\theta + \I (1 - \eta) \sin\theta.\]
The integral~\eqref{eq:Ctaucplx} is hard to compute, but we can make progress by taking the derivative of $C(\tau)$ with respect to $\tau$,
 \begin{equation}
 C'(\tau) = -\frac{\E^{-\tau}}{2\pi g^2} \int_{0}^{2\pi}\int_{0}^{g} \E^{\tau\psi(\theta)r} r\, \d{r}\,\d{\theta},
 \label{eq:Cprime}
 \end{equation}
 which is easier to evaluate. Equation~\eqref{eq:Cprime} can be integrated over $r$ by parts, yielding an integral over $\theta$ only that, excluding prefactors, reads
 \[B(g, \tau) \equiv \int_{0}^{2\pi} \left\{ \frac{g\, \E^{\tau\psi(\theta) g}}{\tau\psi(\theta)} - \frac{\E^{\tau\psi(\theta) g} - 1}{\tau^2 \psi^2(\theta)}\right\} \,\d{\theta}.\]
Again, this integral is hard to compute but we can use the same trick we used before, noting that the partial derivative of $A(g, \tau)$ with respect to $g$ simplifies considerably,
\[\frac{\partial B(g, \tau)}{\partial g} = g \int_{0}^{2\pi} \E^{\tau\psi(\theta) g}\, \d{\theta} = 2 \pi g I_0(2g\tau\sqrt{\eta}),\]
where in the last equation we used \citep[3.937.2, p.~496]{gradshteyn_ryzhik_2007}. We recover the expression for $A(g,\tau)$ by integrating along $g$, with the initial condition $A(0,\tau)=0$:
\[B(g,\tau) = 2\pi \int_{0}^{g} x I_0(2x \tau \sqrt{\eta})\,\d{x} = \frac{2 \pi}{4 \tau^2 \eta} \int_{0}^{2 g \tau \sqrt{\eta}} y I_{0}(y)\,\d{y}. \]
The last integral can be computed with the help of the recurrence relation $zI_0(z) = z I'_1(z) + I_1(z)$. An integration by parts of the term $zI_1'(z)$ leads to the final identity $\int x I_0(x)\, \d{x} = x I_1(x)$ and therefore to
\[B(g,\tau) = \frac{\pi g}{\tau \sqrt{\eta}} I_{1}(2 \tau g \sqrt{\eta}). \]
Equation~\eqref{eq:Cprime} then reads
\begin{equation}
C'(\tau) = -\frac{\E^{-\tau}}{2\tau g \sqrt{\eta}} I_{1}(2 \tau g \sqrt{\eta}),
\label{eq:Cprime_final}
\end{equation}
which we have to integrate to recover $C(\tau)$. Such an integration is subject to the initial condition $C(0)$:
\[\begin{split}
 C(0) &= \frac{1}{2\pi g^2} \int_{0}^{2\pi}\int_{0}^{g} \frac{r}{1 - r \psi(\theta)} \, \d{r}\,\d{\theta}.\\
 &= \frac{-1}{2\pi g^2} \int_{0}^{2\pi}\left\{ \frac{g}{\psi(\theta)} + \frac{1}{\psi^2(\theta)} \ln \bigl[1 - g \psi(\theta)\bigr]\right\}\,\d{\theta},
\end{split}
\]
which can be evaluated numerically.

\section{Summary of the dynamic mean field derivation}
\label{sec:app_DMF}
The starting point of the calculation is the moment generating functional for the state variables $x_i(t)$ obeying Eq.~\eqref{eq:circuit}. We consider the more general case where the activation variable is driven by recurrent inputs as well as independent external white noise:
\begin{equation}
 \dot{x}_i(t) = -x_i(t) + g \sum_{j=1}^{N} J_{ij} r_j(t) + \sigma \xi_i(t),\quad i=1,\dotsc,N\label{eq:dotx}
\end{equation}
where we defined $r_j(t) \equiv \phi\bigl(x_j(t)\bigr)$ to simplify the notation. The white noise sources $\xi_i(t)$ have zero mean and unit variance.
The moment generating functional for such a system can be shown to be~\citetext{\citealp{martin_siggia_rose_pra1973,janssen_zfp1976}. See also \citealp{chow_buice_jmathneuro2015,cugliandolo_lectures2013} for a more pedagogical description}
\begin{multline*}
 Z[l, \tilde{l}; \mathbf{J}] = \int \mathcal{D}x(t) \mathcal{D}\tilde{x}(t) \exp\Bigl(-S[x, \tilde{x}; \mathbf{J}]\\ + \sum_{i=1}^{N}\int \tilde{l}_i(t) x_i(t)\,\d{t} + \sum_{i=1}^{N} \int l_i(t) \tilde{x}_i(t)\,\d{t}, \Bigr)
\end{multline*}
where $\mathcal{D}x(t) \mathcal{D} \tilde{x}(t) = \prod_{i=1}^{N} \mathcal{D}x_i(t) \mathcal{D} \tilde{x}_i(t)$ is the functional measure for all possible paths for all variables and we introduced the action
\begin{multline}
 S[x, \tilde{x}; \mathbf{J}] = \sum_{i=1}^{N} \int \tilde{x}_i(t)\\ \times \biggl\{\dot{x}_i(t) + x_i(t) - g \sum_{j=1}^{N} J_{ij} r_j(t) - \frac{\sigma^2}{2} \tilde{x}_i(t) \biggr\}\,\d{t}
 \label{eq:action}
\end{multline}
In this definition we assume that the auxiliary fields $\tilde{x}(t)$ are purely imaginary, so we do not have to write explicit imaginary units all along. By construction the generating functional satisfies the normalization condition $Z[0,0;\mathbf{J}] = 1$. The fact that $Z[0,0; \mathbf{J}]$ does not depend on $\mathbf{J}$ allows us to compute the quenched average directly on $Z$ \citep{dedominicis_prb1978},
\begin{equation}
 Z[l, \tilde{l}] \equiv \int \frac{Z[l, \tilde{l}; \mathbf{J}]}{Z[0, 0; \mathbf{J}]}\,\d{P(\mathbf{J})} = \int Z[l, \tilde{l}; \mathbf{J}]\, \d{P(\mathbf{J})},
 \label{eq:avgZ_J}
\end{equation}
which simplifies considerably the average, now reduced to computing $[\exp(-S[x,\tilde{x}, \mathbf{J}])]_J$. To do so, we use the decomposition of partially symmetric connectivity matrices
\begin{equation}
 J_{ij} = J_{ij}^{s} + k J_{ij}^{a},
 \label{eq:partial_sym_build}
\end{equation}
where $J_{ij}^{s} = J_{ji}^{s}$, $J_{ij}^{a} = -J_{ji}^{a}$, and where both $J_{ij}^{s}$ and $J_{ij}^{a}$ are Gaussian random variates with zero mean and variance
\[\bigl[ (J_{ij}^{s})^2 \bigr]_{J} = \bigl[ (J_{ij}^{a})^2 \bigr]_{J} = \frac{1}{N}\frac{1}{1 + k^2}\]
so that $[J_{ij}^2]_{J} = J^2 / N$. With these matrix decompositions, the correlation between bidirectional weights is \citep{crisanti_sompolinsky_pra1987}
\[ [J_{ij} J_{ji}]_{J} = \frac{1}{N} \frac{1 - k^2}{1 + k^2},\]
which must equal $\eta / N$ by our definition of $\eta$. This leads to the relation $k^2 = (1 - \eta)/(1 + \eta)$. To integrate over the disorder we use the Gaussian measures:
\begin{align*}
 \d{P}(\mathbf{J}^s) & = \prod_{i \leq j} \d{P}(J_{ij}^s) \propto \exp\biggl\{-\frac{N}{1+\eta} \sum_{i \leq j} (J^s_{ij})^2\biggr\}\d{\mathbf{J}^{s}}, \\
 \d{P}(\mathbf{J}^a) & = \prod_{i < j} \d{P}(J_{ij}^a) \propto \exp\biggl\{-\frac{N}{1+\eta} \sum_{i < j} (J^a_{ij})^2\biggr\}\d{\mathbf{J}^{s}},
\end{align*}
with $\d{\mathbf{J}^{s}} = \prod_{i \leq j} \d{J_{ij}^s}$ and $\d{\mathbf{J}^{s}} = \prod_{i < j} \d{J_{ij}^a}$.
We will ignore the contribution of diagonal elements of the synaptic matrix because it is negligible in the limit of large $N$. We can now integrate out the terms linear in $J_{ij}$ that appear in Eq.~\eqref{eq:action}, by separating symmetric and antisymmetric components. Excluding prefactors and time integrals, these terms are of the form
\begin{multline*}
 L(\mathbf{J}, t) \equiv \sum_{\substack{i, j\\i\neq j}} \tilde{x}_i(t) J_{ij} r_j(t) =
 \sum_{\substack{i, j\\i\neq j}} \tilde{x}_i(t) [J^{s}_{ij} + k J^{a}_{ij}] r_j(t)\\
 = \sum_{i < j} \Bigl\{J^{s}_{ij} \bigl[\tilde{x}_i(t) r_j(t) + \tilde{x}_j(t) r_i(t)\bigr]
 \\ + k J^{a}_{ij} \bigl[\tilde{x}_i(t) r_j(t) - \tilde{x}_j(t) r_i(t)\bigr]\Bigr\}
\end{multline*}
so that
\begin{multline*}
 \int \exp\biggl\{g \int L(\mathbf{J}, t)\,\d{t} \biggr\}\,\d{P}(\mathbf{J}^s)\,\d{P}(\mathbf{J}^a)\\
 = \exp \Biggl\{\frac{g^2}{2N} \sum_{\substack{i, j \\ i \neq j}} \iint \Bigl\{ \bigl[\tilde{x}_i(t) r_j(t) \tilde{x}_i(t') r_j(t')\bigr]\\
 + \eta \bigl[\tilde{x}_i(t) r_j(t) \tilde{x}_j(t') r_i(t')\bigr]\Bigr\}\,\d{t}\,\d{t'} \Biggr\},
\end{multline*}
where we used the property that, for a Gaussian variable $z$ of zero mean and variance $\sigma^{2}$, the expected value of $\exp(\lambda z)$ is $\avg{\exp(\lambda z)}_z = \exp(\lambda^2\sigma^2/2)$, which can be checked by completing the square in the exponential.

Putting back together all the pieces, the average generating functional~\eqref{eq:avgZ_J} is therefore
\begin{multline} Z[l, \tilde{l}] =
 \int \mathcal{D}x(t) \mathcal{D} \tilde{x}(t)\, \exp\Bigl(-S_0[x(t), \tilde{x}(t)]\\ + \frac{\sigma^2}{2} \tilde{x} \cdot \tilde{x} + \tilde{l} \cdot x_i + l \cdot \tilde{x} \\
+ \frac{g^2}{2N} \sum_{\substack{i, j\\ i \neq j}} \iint \Bigl\{ \bigl[\tilde{x}_i(t) r_j(t) \tilde{x}_i(t') r_j(t')\bigr]\\
 + \eta \bigl[\tilde{x}_i(t) r_j(t) \tilde{x}_j(t') r_i(t')\bigr]\Bigr\}\,\d{t}\,\d{t'}
\Bigr)
\label{eq:Zll}
\end{multline}
where we defined the free action
\begin{equation}
 S_0[x, \tilde{x}] \equiv \sum_{i=1}^{N} \int \tilde{x}_i(t)\bigl[\dot{x}_i(t) + x_i(t)\bigr]\,\d{t}.
 \label{eq:free_action}
\end{equation}
and we introduced the notation
\[f \cdot g \equiv \sum_{i=1}^{N}\int f_i(t) g_i(t)\,\d{t}.\]

As a result of averaging out the disorder, we obtained a coupling involving four fields with different indices and at different times. To proceed it is convenient to introduce auxiliary fields that involve terms local in space (i.e., with the same index) but not in time:
\begin{align*}
 q_{1}(t, t') &= \frac{g^2}{N} \sum_{i=1}^{N} \tilde{x}_i(t) \tilde{x}_i(t'), &
 q_{2}(t, t') &= \frac{g^2}{N} \sum_{i=1}^{N} r_i(t) r_i(t'),\\
 q_{3}(t, t') &= \frac{g^2}{N} \sum_{i=1}^{N} \tilde{x}_i(t) r_i(t'), &
 q_{4}(t, t') &= \frac{g^2}{N} \sum_{i=1}^{N} r_i(t) \tilde{x}_i(t'),
\end{align*}
so Eq.~\eqref{eq:Zll} now reads
\begin{multline}
Z[l, \tilde{l}] =
\int \mathcal{D}x(t) \mathcal{D} \tilde{x}(t) \left(\prod_{\alpha=1}^{4} \frac{N}{g^2}\, \mathcal{D}q_{\alpha}\right)\\ \times \delta\biggl(\frac{N}{g^{2}} q_1 - \sum_{i=1}^{N} \tilde{x}_i(t) \tilde{x}_i(t')\biggr) \delta\biggl(\frac{N}{g^{2}} q_2 - \sum_{i=1}^{N} r_i(t) r_i(t')\biggr) \\
\times \delta\biggl(\frac{N}{g^{2}} q_3 - \sum_{i=1}^{N} \tilde{x}_i(t) r_i(t')\biggr) \delta\biggl(\frac{N}{g^{2}} q_4 - \sum_{i=1}^{N} r_i(t) \tilde{x}_i(t')\biggr)\\
\times \exp\biggl(-S_0[x, \tilde{x}] + \frac{\sigma^2}{2} \tilde{x} \cdot \tilde{x} + \tilde{l} \cdot x_i + l \cdot \tilde{x} \\ 
+ \frac{N}{2 g^2} \iint \biggl\{ q_1(t, t') q_2(t, t') + \eta\, q_3(t, t') q_4(t, t') \biggr\}\,\d{t}\,\d{t'} \biggr)
\label{eq:Zll_deltas}
\end{multline}
We now express the Dirac functionals in their integral representation. The first Dirac functional appearing in Eq.~\eqref{eq:Zll_deltas} can be written as
\begin{multline*}
 \delta\biggl(\frac{N}{g^{2}} q_1 - \sum_{i=1}^{N} \tilde{x}_i(t) \tilde{x}_i(t')\biggr) = \frac{1}{2\pi} \int \mathcal{D}\hat{q}_1(t, t') \\ \times \exp\biggl\{ \iint \hat{q}_1(t, t') \Bigl[\frac{N}{g^{2}} q_1(t, t') - \sum_{i=1}^{N} \tilde{x}_i(t) \tilde{x}_i(t') \Bigr]\,\d{t}\,\d{t}' \biggr\},
\end{multline*}
where the integral over $\hat{q}$ is understood to be along the imaginary axis. The other Dirac functionals in Eq.~\eqref{eq:Zll_deltas} are rewritten analogously. Equation~\eqref{eq:Zll_deltas} then becomes
\begin{multline}
Z[l, \tilde{l}] =
\int \mathcal{D}X \mathcal{D}Q\, \exp\Bigl(-S_0[x, \tilde{x}] + \frac{\sigma^2}{2}\, \tilde{x}\cdot \tilde{x} + \tilde{l} \cdot x + l \cdot \tilde{x}\\
+ \frac{N}{g^2} \iint \biggl\{ \sum_{\alpha = 1}^{4} \hat{q}_{\alpha}(t, t') q_{\alpha}(t, t')\\
+ \frac{1}{2} \bigl[ q_1(t, t') q_2(t, t') + \eta\, q_3(t, t') q_4(t, t')\bigr] \\
- \frac{g^2}{N} \sum_{i=1}^{N} \Bigl[ \hat{q}_1(t, t') \tilde{x}_i(t) \tilde{x}_i(t') +
\hat{q}_2(t, t') r_i(t) r_i(t')\\ + \hat{q}_3(t, t') \tilde{x}_i(t) r_i(t') + \hat{q}_4(t, t') r_i(t) \tilde{x}_i(t')\Bigr]\biggr\}\,\d{t}\,\d{t'}
\biggr)
\label{eq:Zaux}
\end{multline}
where we introduced the shorthand notation
\begin{align*}
 \mathcal{D} Q & \equiv \prod_{\alpha=1}^{4} \frac{1}{2\pi} \frac{N}{g^2} \mathcal{D} q_{\alpha} \mathcal{D}\hat{q}_{\alpha},\\
 \mathcal{D} X & \equiv \mathcal{D}x(t) \mathcal{D} \tilde{x}(t) = \prod_{i=1}^{N} \mathcal{D}x_i(t) \mathcal{D} \tilde{x}_i(t).
\end{align*}

Equation~\eqref{eq:Zaux} can now be expressed as~\citep{sompolinsky_zippelius_prb1982,toyoizumi_abbott_pre2011}
\begin{equation}
 Z[l, \tilde{l}] = \int \mathcal{D}Q\, \E^{N f(q, \hat{q}, x, \tilde{x})},
 \label{eq:ZexpNf}
\end{equation}
where
\begin{align*}
 f(q, \hat{q}, x, \tilde{x}) & \equiv G(q, \hat{q}) + \frac{1}{N} \log \int \mathcal{D}X\, \exp\bigl[\mathcal{L}(q, \hat{q}, x, \tilde{x})\bigr],\\
 G(q, \hat{q}) & \equiv \frac{1}{g^2} \iint \Bigl\{ \sum_{\alpha=1}^{4} q_{\alpha} \hat{q}_{\alpha} + \frac{1}{2} \bigl[ q_1 q_2 + \eta\, q_3 q_4 \bigr] \Bigr\}\,\d{t}\,\d{t'},\\
 \mathcal{L}(q, \hat{q}, x, \tilde{x}) & \equiv -S_0[x, \tilde{x}] + \frac{\sigma^2}{2}\, \tilde{x} \cdot \tilde{x} + \tilde{l} \cdot x + l \cdot \tilde{x} \\
 & - \sum_{i=1}^{N} \iint \bigl[ \hat{q}_1(t, t') \tilde{x}_i(t) \tilde{x}_i(t') +
\hat{q}_2(t, t') r_i(t) r_i(t') \\
& \hspace*{1em} + \hat{q}_3(t, t') \tilde{x}_i(t) r_i(t') + \hat{q}_4(t, t') r_i(t) \tilde{x}_i(t')\bigr]\,\d{t}\,\d{t'}.
\end{align*}
In the limit of large $N$ we can apply the saddle-point method to Eq.~\eqref{eq:ZexpNf}, which amounts to making the following approximation
 \begin{equation}
 Z[l, \tilde{l}] = \int \mathcal{D}Q\, \E^{N f(q, \hat{q}, x, \tilde{x})} \approx \E^{N f(q^0\!, \hat{q}^0\!, x, \tilde{x})}
 \label{eq:Zll_noDX}
 \end{equation}
 where $q^0$ and $\hat{q}^0$ are the values that extremize $f$. Requiring $\delta f/\delta \hat{q}_{\alpha} = 0$ leads to
 \begin{align}
 q^{0}_{1}(t, t') &= \frac{g^2}{N} \sum_{i=1}^{N} \bigl\langle \tilde{x}_i(t) \tilde{x}_i(t')\bigr\rangle_{\!\mathcal{L}}\:, \label{eq:q12}\\ 
 q^{0}_{2}(t, t') &= \frac{g^2}{N} \sum_{i=1}^{N} \bigl\langle r_i(t) r_i(t')\bigr\rangle_{\!\mathcal{L}}\:,\\
 q^{0}_{3}(t, t') &= \frac{g^2}{N} \sum_{i=1}^{N} \bigl\langle \tilde{x}_i(t) r_i(t')\bigr\rangle_{\!\mathcal{L}}\:,\\
 q^{0}_{4}(t, t') &= \frac{g^2}{N} \sum_{i=1}^{N} \bigl\langle r_i(t) \tilde{x}_i(t')\bigr\rangle_{\!\mathcal{L}}\:, \label{eq:q34}
 \end{align}
 with the average $\langle \cdot \rangle_{\mathcal{L}}$ defined as
\[ \avg{\mathcal{O}}_{\mathcal{L}} \equiv \dfrac{\int \mathcal{O}(X)\, \exp\bigl[\mathcal{L}(X)\bigr]\,\d{X}}{\int \exp\bigl[\mathcal{L}(X)\bigr]\,\d{X}}.\]
Similarly, from the saddle-point conditions for $q_{\alpha}$ we obtain
\begin{align*}
 \hat{q}^{0}_{1}(t, t') &= -\frac{1}{2} q^{0}_{2}(t, t'), &
 \hat{q}^{0}_{2}(t, t') &= -\frac{1}{2}q^{0}_{1}(t, t'),\\
 \hat{q}^{0}_{3}(t, t') &= - \frac{\eta}{2} q^{0}_{4}(t, t'), &
 \hat{q}^{0}_{4}(t, t') &= - \frac{\eta}{2} q^{0}_{3}(t, t').
\end{align*}

Now the right hand side of Eq.~\eqref{eq:Zll_noDX} reads
\begin{equation}
Z[l, \tilde{l}] =
\int \mathcal{D}X\, \exp\Bigl(-S_0[x, \tilde{x}] - S_{\text{int}}[x, \tilde{x}] + \tilde{l} \cdot x + l \cdot \tilde{x} \Bigr)
\label{eq:Zll_eff}
\end{equation}
with
\begin{multline}
S_{\text{int}}[x, \tilde{x}] \equiv
\frac{1}{2} \iint \biggl\{\frac{N}{g^2}\Bigl[q^{0}_1(t, t') q^{0}_2(t, t') 
\eta \, q^{0}_3(t, t') q^{0}_4(t, t') \Bigr]
\\ - \sum_{i=1}^{N} \Bigl\{\bigl[q^{0}_2(t, t') + \sigma^2 \delta(t - t')\bigr]\tilde{x}_i(t) \tilde{x}_i(t')\\
+ q^{0}_1(t, t') r_i(t) r_i(t') \eta \, q^{0}_4(t, t') \tilde{x}_i(t) r_i(t')\\ \eta \, q^{0}_3(t, t') r_i(t) \tilde{x}_i(t')\Bigr\}\biggr\}\,\d{t}\,\d{t'}.
\label{eq:S_int}
\end{multline}

The auxiliary fields defined in Eqs.~\eqref{eq:q12}--\eqref{eq:q34} are related to physically observable quantities.
 First, $q_2^0(t, t')$ is related to the population-averaged autocorrelation function
\[ C(t, t') \equiv \frac{1}{N} \sum_{i=1}^{N} \bigl\langle r_i(t) r_i(t') \bigr\rangle\:,\]
by $q_2^{0} (t, t') = g^2 C(t, t')$.

Second, the auxiliary fields $q_3^0(t, t')$ and $q_4^{0}(t, t')$ are related to the so-called response function, which characterizes the response of the system when it is perturbed by a weak field. More specifically, in our context the response function at site $i$ would be
\begin{equation}
 G(t, t') \equiv \left. \frac{\delta \langle r_i(t) \rangle}{\delta h_i(t')} \right|_{h=0},
 \label{eq:defG}
\end{equation}
where $h_i(t')$ is a time-dependent external field, and angular brackets denote the average over the effective action $S[x,\tilde{x}] = S_0[x,\tilde{x}] + S_{\text{int}}[x,\tilde{x}]$ that appears in Eq.~\eqref{eq:Zll_eff}. Note that from the definition of response function $G(t, t')$ has to be 0 whenever $t < t'$, due to causality. To see the link between $G(t, t')$ and $q_3(t,t')$ and $q_4(t,t')$, we add an external field $h_i(t)$ for each neuron in Eq.~\eqref{eq:dotx}, and evaluate \eqref{eq:defG}. With the new field the action becomes $S_h[x,\tilde{x}] = S[x,\tilde{x}] -\sum \tilde{x}_i(t) h_i(t)$ and
\[\begin{split} \left. \frac{\delta \langle r_i(t) \rangle}{\delta h_i(t')} \right|_{h=0}
 & = \frac{\delta}{\delta h(t')} \int \mathcal{D}X\, r_i(t) \exp\bigl(-S_h[x, \tilde{x}]\bigr) \biggl|_{h=0} \\ 
& = -\left \langle r(t) \left. \frac{\delta S_h}{\delta h_i(t')}\right|_{h=0} \right\rangle = \langle r_i(t) \tilde{x}_i(t') \rangle.
\end{split}\]
Defining the population-averaged response function as
\[ G(t, t') \equiv \frac{1}{N} \sum_{i=1}^{N} \bigl\langle r_i(t) \tilde{x}_i(t') \bigr\rangle\:,\]
we obtain $q_4^{0}(t, t') = q_3^{0}(t', t) = g^2 G(t, t')$.

As for $q_1^{0}(t,t')$, it can be shown that the presence of vertices like
$r_i(t) r_i(t')$ in the action necessarily leads to violation of causality
\citep{sompolinsky_zippelius_prb1982}. We thus need to impose $q_1^{0}(t,t')=0$
to obtain a physical solution.

We can finally write the interacting action in Eq.~\eqref{eq:S_int} in terms of the physical quantities
\begin{multline}
 S_{\text{int}}[x,\tilde{x}] = -\sum_{i=1}^{N} \iint \biggl\{ \frac{1}{2} \Gamma(t, t') \tilde{x}_i(t) \tilde{x}_i(t') \\+ \eta g^2\, G(t, t') \tilde{x}_i(t) r_i(t') \biggr\}\,\d{t}\,\d{t'}, \label{eq:inter_action}
\end{multline}
where we ignored the term containing $G(t, t')G(t, t')$, which vanishes due to causality, and where we defined
\[\Gamma(t, t') \equiv g^2 C(t, t') + \sigma^2 \delta(t - t')\:.\]
Note that the final action involves only interactions that are local in space, which implies that all units are equivalent. This equivalence comes as no surprise, because all units are equivalent once we average over all realizations of the connectivity matrix. We can thus drop the irrelevant indices $i$ and focus on the single relevant dynamical variable $x(t)$.

\subsection*{Equation of motion for the average activity}
The local action $S[x, \tilde{x}] = S_0[x, \tilde{x}] +S_{\text{int}}[x, \tilde{x}]$, with $S_0$ and $S_{\text{int}}$ given by
Eqs.~\eqref{eq:free_action} and \eqref{eq:inter_action}, has the form
\begin{multline}
 S[x, \tilde{x}] = \iint \Bigl\{\tilde{x}(t) G^{-1}_{F}(t - t') x(t') - \eta g^2 \tilde{x}(t) G(t, t') r(t') \\
 - \frac{1}{2}\Gamma(t, t') \tilde{x}(t)\tilde{x}(t')\Bigr\}\, \d{t}\,\d{t'}
\label{eq:eff_action}
\end{multline}
where for later convenience we have introduced the inverse of the free propagator, $G^{-1}_F(t - t')$. The free propagator $G_F(t - t')$ is just the Green's function associated with the operator $\d{}/\d{t} + 1$, \[\biggl(\frac{\d{}}{\d{t}} + 1\biggr) G_{F}(t - t') = \delta(t - t'),\]
and is related to its inverse through $\int G_F^{-1}(t - s)G_F(s - t')\,\d{s} = \delta( t - t')$, which is automatically satisfied if
\[ G_F^{-1} (t - t') = \delta(t - t') \biggl(\frac{\d{}}{\d{t'}} + 1\biggr). \]

From the original stochastic system~\eqref{eq:dotx} and its associated
Martin-Siggia-Rose-Janssen-de Dominicis (MSRJD)
action~\eqref{eq:action}, we infer that the equation of motion
associated with the action~\eqref{eq:eff_action} is
\begin{equation}
\dot{x}(t) = - x(t) + \eta g^2 \int_{-\infty}^{t} G(t, s) r(s)\,\d{s} + \varphi(t),
\label{eq:motion}
\end{equation}
where $\varphi(t)$ is a source of noise with autocorrelation
\[\langle \varphi(t) \varphi(t') \rangle = \Gamma(t, t') = g^2 C(t, t') + \sigma^2 \delta(t - t').\]
This relation has to be consistent with the dynamics generated by Eq.~\eqref{eq:motion}, that is, the noise $\varphi(t)$ has to be such that the firing activity $r(t)$ has autocorrelation $C(t, t')$.

We can go further and write a self-consistent relation involving the two-point functions $C(t,t')$ and $G(t, t')$. A starting point to derive them are the identities
\begin{equation*}
 \frac{\delta x(t)}{\delta x(t')} = \frac{\delta \tilde{x}(t)}{\delta \tilde{x}(t')} = \delta(t - t'),\quad
\frac{\delta \tilde{x}(t)}{\delta x(t')} = \frac{\delta x(t)}{\delta \tilde{x}(t')} = 0
\end{equation*}
from which we can obtain relations such as
\begin{align*}
 \left \langle \frac{\delta x(t)}{\delta x(t')} \right\rangle
 & \equiv \int \mathcal{D}X\, \frac{\delta x(t)}{\delta x(t')} \exp\bigl\{-S[x(t), \tilde{x}(t)]\bigr\} \\ & = \left \langle x(t) \frac{\delta S}{\delta x(t')} \right\rangle = \delta(t - t').
\end{align*}
Other relations follow analogously:
\begin{align}
 \left \langle x(t) \frac{\delta S}{\delta x(t')} \right\rangle & = \delta(t - t'), &
 \left \langle x(t) \frac{\delta S}{\delta \tilde{x}(t')} \right\rangle & = 0,\label{eq:xS1}\\
 \left \langle \tilde{x}(t) \frac{\delta S}{\delta \tilde{x}(t')} \right\rangle & = \delta(t - t'),
 & \left \langle \tilde{x}(t) \frac{\delta S}{\delta x(t')} \right\rangle & = 0,\label{eq:xS2}
\end{align}

We now apply the identities~\eqref{eq:xS1} and \eqref{eq:xS2} for the action~\eqref{eq:eff_action}. In particular, we use the identities involving
\[\frac{\delta S}{\delta \tilde{x}(t')} = \dot{x}(t) + x(t) - \eta g^2 \! \int_{-\infty}^{t}\!\! G(t,s) r(s) \,\d{s} - \int \Gamma(t, s) \tilde{x}(s)\,\d{s}\]
and we define the autocorrelation and response function of the activation field $x(t)$
\begin{equation*}
 \Delta(t, t') \equiv \avg{x(t) x(t')},\quad R(t, t') \equiv \avg{x(t) \tilde{x}(t')}.
\end{equation*}
The last equation in \eqref{eq:xS1} and the first equation in \eqref{eq:xS2} then become, respectively,
 \begin{align}
 \frac{\partial}{\partial t} \Delta(t, t') & = -\Delta(t, t') + \sigma^2 R(t', t)
 \eta g^2 \!\! \int^{t}_{0} \!\! G(t, s) \avg{r(s)x(t')}\,\d{s}\notag \\
 & \qquad + g^2 \int^{t'}_{0} R(t',s) C(t,s)\,\d{s},\label{eq:SD1}\\
 \frac{\partial}{\partial t} R(t, t') & = -R(t, t') + \delta(t - t')\notag \\
 & \qquad + \eta g^2 \int^{t}_{t'} G(t, s) G(s,t')\,\d{s}\label{eq:SD2},
\end{align}
where in~\eqref{eq:SD2} we have used $\avg{\tilde{x}(t)\tilde{x}(t')} = 0$. It can be shown that the remaining identities in Eqs.~\eqref{eq:xS1} and \eqref{eq:xS2}, which involve $\delta S / \delta x$, do not provide additional information~\citep{cugliandolo_lectures2013}. Note that $\Delta(t,t')$ has a cusp at $t=t'$ due to the term $\sigma^2 R(t', t)$, which from \eqref{eq:xS2} we know it must be of the form $R(t,t') \propto \Theta(t - t')$, with $\Theta(t)$ being the step function. More specifically,
\begin{equation*}
 \left[\frac{\partial}{\partial t} \Delta(t, t') \right]_{t'= t^{+}}^{t'= t^{-}} = \sigma^2 \bigl[R(t',t)\bigr]_{t'= t^{+}}^{t'= t^{-}} = -\sigma^2.
\end{equation*}
Moreover, the symmetry of $\Delta(t, t')$ around $t=t'$ implies $\lim_{t'\rightarrow t^{-}} \partial_t \Delta(t, t') = - \lim_{t'\rightarrow t^{+}} \partial_t \Delta(t, t')$, which leads to the relation $\lim_{t'\rightarrow t^{-}} \partial_t \Delta(t, t') = -\sigma^2 / 2$. The amplitude of external noise thus determines the slope of the autocorrelation of $x(t)$ at zero time lag. This is the only dependence on $\sigma^2$ of the solutions of~\eqref{eq:SD1} and \eqref{eq:SD2}.

Equations~\eqref{eq:SD1} and \eqref{eq:SD2} cannot be solved in a closed-form except for $\eta=0$ \citep{sompolinsky_crisanti_sommers_prl1988}, but perturbative solutions can be found by expanding the nonlinearity $r(t)=\phi(x)$ in power series of $x(t)$ and then solving the resulting hierarchy of equations, which involve correlations and response functions of increasingly larger order. The problem becomes unwieldly except for the linear case where $r(t)=x(t)$. In that case, $C(t,t')=\Delta(t,t')$, $G(t,t')=R(t, t')$, and Eqs.~\eqref{eq:SD1} and \eqref{eq:SD2} form a closed system of integro-differential equations:
\begin{align}
\frac{\partial}{\partial t} \Delta(t, t') & = -\Delta(t, t') + \sigma^2 R(t', t) +
\eta J^2 \!\! \int^{t}_{0} \!\! R(t, s) \Delta(s, t')\,\d{s} \notag \\
& \hspace{2cm} + g^2 \int^{t'}_{0} R(t',s) \Delta(t,s)\,\d{s},\label{eq:SD1lin}\\
\frac{\partial}{\partial t} R(t, t') & = -R(t, t') + \delta(t - t')\notag \\ & \hspace{2cm} + \eta g^2 \int^{t}_{t'} R(t, s) R(s, t')\,\d{s}.\label{eq:SD2lin}
\end{align}

\bibliographystyle{apsrev4-1}
\bibliography{ms}

\begin{thebibliography}{74}%
\makeatletter
\providecommand \@ifxundefined [1]{%
 \@ifx{#1\undefined}
}%
\providecommand \@ifnum [1]{%
 \ifnum #1\expandafter \@firstoftwo
 \else \expandafter \@secondoftwo
 \fi
}%
\providecommand \@ifx [1]{%
 \ifx #1\expandafter \@firstoftwo
 \else \expandafter \@secondoftwo
 \fi
}%
\providecommand \natexlab [1]{#1}%
\providecommand \enquote  [1]{``#1''}%
\providecommand \bibnamefont  [1]{#1}%
\providecommand \bibfnamefont [1]{#1}%
\providecommand \citenamefont [1]{#1}%
\providecommand \href@noop [0]{\@secondoftwo}%
\providecommand \href [0]{\begingroup \@sanitize@url \@href}%
\providecommand \@href[1]{\@@startlink{#1}\@@href}%
\providecommand \@@href[1]{\endgroup#1\@@endlink}%
\providecommand \@sanitize@url [0]{\catcode `\\12\catcode `\$12\catcode
  `\&12\catcode `\#12\catcode `\^12\catcode `\_12\catcode `\%12\relax}%
\providecommand \@@startlink[1]{}%
\providecommand \@@endlink[0]{}%
\providecommand \url  [0]{\begingroup\@sanitize@url \@url }%
\providecommand \@url [1]{\endgroup\@href {#1}{\urlprefix }}%
\providecommand \urlprefix  [0]{URL }%
\providecommand \Eprint [0]{\href }%
\providecommand \doibase [0]{http://dx.doi.org/}%
\providecommand \selectlanguage [0]{\@gobble}%
\providecommand \bibinfo  [0]{\@secondoftwo}%
\providecommand \bibfield  [0]{\@secondoftwo}%
\providecommand \translation [1]{[#1]}%
\providecommand \BibitemOpen [0]{}%
\providecommand \bibitemStop [0]{}%
\providecommand \bibitemNoStop [0]{.\EOS\space}%
\providecommand \EOS [0]{\spacefactor3000\relax}%
\providecommand \BibitemShut  [1]{\csname bibitem#1\endcsname}%
\let\auto@bib@innerbib\@empty
\bibitem [{\citenamefont {Sompolinsky}\ \emph {et~al.}(1988)\citenamefont
  {Sompolinsky}, \citenamefont {Crisanti},\ and\ \citenamefont
  {Sommers}}]{sompolinsky_crisanti_sommers_prl1988}%
  \BibitemOpen
  \bibfield  {author} {\bibinfo {author} {\bibfnamefont {H.}~\bibnamefont
  {Sompolinsky}}, \bibinfo {author} {\bibfnamefont {A.}~\bibnamefont
  {Crisanti}}, \ and\ \bibinfo {author} {\bibfnamefont {H.~J.}\ \bibnamefont
  {Sommers}},\ }\href {\doibase 10.1103/PhysRevLett.61.259} {\bibfield
  {journal} {\bibinfo  {journal} {Phys. Rev. Lett.}\ }\textbf {\bibinfo
  {volume} {61}},\ \bibinfo {pages} {259} (\bibinfo {year} {1988})}\BibitemShut
  {NoStop}%
\bibitem [{\citenamefont {van Vreeswijk}\ and\ \citenamefont
  {Sompolinsky}(1998)}]{vanvreeswijk_sompolinsky_neco1998}%
  \BibitemOpen
  \bibfield  {author} {\bibinfo {author} {\bibfnamefont {C.}~\bibnamefont {van
  Vreeswijk}}\ and\ \bibinfo {author} {\bibfnamefont {H.}~\bibnamefont
  {Sompolinsky}},\ }\href {\doibase 10.1162/089976698300017214} {\bibfield
  {journal} {\bibinfo  {journal} {Neural Comput.}\ }\textbf {\bibinfo {volume}
  {10}},\ \bibinfo {pages} {1321} (\bibinfo {year} {1998})}\BibitemShut
  {NoStop}%
\bibitem [{\citenamefont {Brunel}(2000)}]{brunel_jcns2000}%
  \BibitemOpen
  \bibfield  {author} {\bibinfo {author} {\bibfnamefont {N.}~\bibnamefont
  {Brunel}},\ }\href {\doibase 10.1023/a:1008925309027} {\bibfield  {journal}
  {\bibinfo  {journal} {J. Comput. Neurosci.}\ }\textbf {\bibinfo {volume}
  {8}},\ \bibinfo {pages} {183} (\bibinfo {year} {2000})}\BibitemShut {NoStop}%
\bibitem [{\citenamefont {Buonomano}\ and\ \citenamefont
  {Maass}(2009)}]{buonomano_maass_natns2009}%
  \BibitemOpen
  \bibfield  {author} {\bibinfo {author} {\bibfnamefont {D.~V.}\ \bibnamefont
  {Buonomano}}\ and\ \bibinfo {author} {\bibfnamefont {W.}~\bibnamefont
  {Maass}},\ }\href {\doibase 10.1038/nrn2558} {\bibfield  {journal} {\bibinfo
  {journal} {Nat. Rev. Neurosci.}\ }\textbf {\bibinfo {volume} {10}},\ \bibinfo
  {pages} {113} (\bibinfo {year} {2009})}\BibitemShut {NoStop}%
\bibitem [{\citenamefont {Toyoizumi}\ and\ \citenamefont
  {Abbott}(2011)}]{toyoizumi_abbott_pre2011}%
  \BibitemOpen
  \bibfield  {author} {\bibinfo {author} {\bibfnamefont {T.}~\bibnamefont
  {Toyoizumi}}\ and\ \bibinfo {author} {\bibfnamefont {L.~F.}\ \bibnamefont
  {Abbott}},\ }\href {\doibase 10.1103/PhysRevE.84.051908} {\bibfield
  {journal} {\bibinfo  {journal} {Phys. Rev. E}\ }\textbf {\bibinfo {volume}
  {84}},\ \bibinfo {pages} {051908} (\bibinfo {year} {2011})}\BibitemShut
  {NoStop}%
\bibitem [{\citenamefont {Sussillo}\ and\ \citenamefont
  {Abbott}(2009)}]{sussillo_abbott_neuron2009}%
  \BibitemOpen
  \bibfield  {author} {\bibinfo {author} {\bibfnamefont {D.}~\bibnamefont
  {Sussillo}}\ and\ \bibinfo {author} {\bibfnamefont {L.~F.}\ \bibnamefont
  {Abbott}},\ }\href {\doibase 10.1016/j.neuron.2009.07.018} {\bibfield
  {journal} {\bibinfo  {journal} {Neuron}\ }\textbf {\bibinfo {volume} {63}},\
  \bibinfo {pages} {544} (\bibinfo {year} {2009})}\BibitemShut {NoStop}%
\bibitem [{\citenamefont {Shadlen}\ and\ \citenamefont
  {Newsome}(1998)}]{shadlen_newsome_jns1998}%
  \BibitemOpen
  \bibfield  {author} {\bibinfo {author} {\bibfnamefont {M.~N.}\ \bibnamefont
  {Shadlen}}\ and\ \bibinfo {author} {\bibfnamefont {W.~T.}\ \bibnamefont
  {Newsome}},\ }\href@noop {} {\bibfield  {journal} {\bibinfo  {journal} {J.
  Neurosci.}\ }\textbf {\bibinfo {volume} {18}},\ \bibinfo {pages} {3870}
  (\bibinfo {year} {1998})}\BibitemShut {NoStop}%
\bibitem [{\citenamefont {Amit}\ and\ \citenamefont
  {Brunel}(1997)}]{amit_brunel_cercor1997}%
  \BibitemOpen
  \bibfield  {author} {\bibinfo {author} {\bibfnamefont {D.~J.}\ \bibnamefont
  {Amit}}\ and\ \bibinfo {author} {\bibfnamefont {N.}~\bibnamefont {Brunel}},\
  }\href {\doibase 10.1093/cercor/7.3.237} {\bibfield  {journal} {\bibinfo
  {journal} {Cereb. Cortex}\ }\textbf {\bibinfo {volume} {7}},\ \bibinfo
  {pages} {237} (\bibinfo {year} {1997})}\BibitemShut {NoStop}%
\bibitem [{\citenamefont {Markram}\ \emph {et~al.}(1997)\citenamefont
  {Markram}, \citenamefont {L{\"u}bke}, \citenamefont {Frotscher},\ and\
  \citenamefont {Sakmann}}]{markram_etal_jphysiol1997}%
  \BibitemOpen
  \bibfield  {author} {\bibinfo {author} {\bibfnamefont {H.}~\bibnamefont
  {Markram}}, \bibinfo {author} {\bibfnamefont {J.}~\bibnamefont {L{\"u}bke}},
  \bibinfo {author} {\bibfnamefont {M.}~\bibnamefont {Frotscher}}, \ and\
  \bibinfo {author} {\bibfnamefont {B.}~\bibnamefont {Sakmann}},\ }\href@noop
  {} {\bibfield  {journal} {\bibinfo  {journal} {J. Physiol.}\ }\textbf
  {\bibinfo {volume} {500}},\ \bibinfo {pages} {409} (\bibinfo {year}
  {1997})}\BibitemShut {NoStop}%
\bibitem [{\citenamefont {Song}\ \emph {et~al.}(2005)\citenamefont {Song},
  \citenamefont {Sj{\"o}str{\"o}m}, \citenamefont {Reigl}, \citenamefont
  {Nelson},\ and\ \citenamefont {Chklovskii}}]{song_etal_plosbiol2005}%
  \BibitemOpen
  \bibfield  {author} {\bibinfo {author} {\bibfnamefont {S.}~\bibnamefont
  {Song}}, \bibinfo {author} {\bibfnamefont {P.~J.}\ \bibnamefont
  {Sj{\"o}str{\"o}m}}, \bibinfo {author} {\bibfnamefont {M.}~\bibnamefont
  {Reigl}}, \bibinfo {author} {\bibfnamefont {S.}~\bibnamefont {Nelson}}, \
  and\ \bibinfo {author} {\bibfnamefont {D.~B.}\ \bibnamefont {Chklovskii}},\
  }\href {\doibase 10.1371/journal.pbio.0030068} {\bibfield  {journal}
  {\bibinfo  {journal} {{PLoS} Biol.}\ }\textbf {\bibinfo {volume} {3}},\
  \bibinfo {pages} {e68} (\bibinfo {year} {2005})}\BibitemShut {NoStop}%
\bibitem [{\citenamefont {Perin}\ \emph {et~al.}(2011)\citenamefont {Perin},
  \citenamefont {Berger},\ and\ \citenamefont {Markram}}]{perin_etal_pnas2011}%
  \BibitemOpen
  \bibfield  {author} {\bibinfo {author} {\bibfnamefont {R.}~\bibnamefont
  {Perin}}, \bibinfo {author} {\bibfnamefont {T.~K.}\ \bibnamefont {Berger}}, \
  and\ \bibinfo {author} {\bibfnamefont {H.}~\bibnamefont {Markram}},\ }\href
  {\doibase 10.1073/pnas.1016051108} {\bibfield  {journal} {\bibinfo  {journal}
  {P. Natl. Acad. Sci. USA}\ }\textbf {\bibinfo {volume} {108}},\ \bibinfo
  {pages} {5419} (\bibinfo {year} {2011})}\BibitemShut {NoStop}%
\bibitem [{\citenamefont {Ko}\ \emph {et~al.}(2011)\citenamefont {Ko},
  \citenamefont {Hofer}, \citenamefont {Pichler}, \citenamefont {Buchanan},
  \citenamefont {Sj{\"o}str{\"o}m},\ and\ \citenamefont
  {Mrsic-Flogel}}]{ko_etal_nature2011}%
  \BibitemOpen
  \bibfield  {author} {\bibinfo {author} {\bibfnamefont {H.}~\bibnamefont
  {Ko}}, \bibinfo {author} {\bibfnamefont {S.~B.}\ \bibnamefont {Hofer}},
  \bibinfo {author} {\bibfnamefont {B.}~\bibnamefont {Pichler}}, \bibinfo
  {author} {\bibfnamefont {K.~A.}\ \bibnamefont {Buchanan}}, \bibinfo {author}
  {\bibfnamefont {P.~J.}\ \bibnamefont {Sj{\"o}str{\"o}m}}, \ and\ \bibinfo
  {author} {\bibfnamefont {T.~D.}\ \bibnamefont {Mrsic-Flogel}},\ }\href
  {\doibase 10.1038/nature09880} {\bibfield  {journal} {\bibinfo  {journal}
  {Nature}\ }\textbf {\bibinfo {volume} {473}},\ \bibinfo {pages} {87}
  (\bibinfo {year} {2011})}\BibitemShut {NoStop}%
\bibitem [{\citenamefont {Harris}\ and\ \citenamefont
  {Mrsic-Flogel}(2013)}]{harris_mrsic-flogel_nature2013}%
  \BibitemOpen
  \bibfield  {author} {\bibinfo {author} {\bibfnamefont {K.~D.}\ \bibnamefont
  {Harris}}\ and\ \bibinfo {author} {\bibfnamefont {T.~D.}\ \bibnamefont
  {Mrsic-Flogel}},\ }\href {\doibase 10.1038/nature12654} {\bibfield  {journal}
  {\bibinfo  {journal} {Nature}\ }\textbf {\bibinfo {volume} {503}},\ \bibinfo
  {pages} {51} (\bibinfo {year} {2013})}\BibitemShut {NoStop}%
\bibitem [{\citenamefont {Wang}\ \emph {et~al.}(2006)\citenamefont {Wang},
  \citenamefont {Markram}, \citenamefont {Goodman}, \citenamefont {Berger},
  \citenamefont {Ma},\ and\ \citenamefont
  {Goldman-Rakic}}]{wang_etal_natns2006}%
  \BibitemOpen
  \bibfield  {author} {\bibinfo {author} {\bibfnamefont {Y.}~\bibnamefont
  {Wang}}, \bibinfo {author} {\bibfnamefont {H.}~\bibnamefont {Markram}},
  \bibinfo {author} {\bibfnamefont {P.~H.}\ \bibnamefont {Goodman}}, \bibinfo
  {author} {\bibfnamefont {T.~K.}\ \bibnamefont {Berger}}, \bibinfo {author}
  {\bibfnamefont {J.}~\bibnamefont {Ma}}, \ and\ \bibinfo {author}
  {\bibfnamefont {P.~S.}\ \bibnamefont {Goldman-Rakic}},\ }\href {\doibase
  10.1038/nn1670} {\bibfield  {journal} {\bibinfo  {journal} {Nat. Neurosci.}\
  }\textbf {\bibinfo {volume} {9}},\ \bibinfo {pages} {534} (\bibinfo {year}
  {2006})}\BibitemShut {NoStop}%
\bibitem [{\citenamefont {Crisanti}\ and\ \citenamefont
  {Sompolinsky}(1987)}]{crisanti_sompolinsky_pra1987}%
  \BibitemOpen
  \bibfield  {author} {\bibinfo {author} {\bibfnamefont {A.}~\bibnamefont
  {Crisanti}}\ and\ \bibinfo {author} {\bibfnamefont {H.}~\bibnamefont
  {Sompolinsky}},\ }\href {\doibase 10.1103/PhysRevA.36.4922} {\bibfield
  {journal} {\bibinfo  {journal} {Phys. Rev. A}\ }\textbf {\bibinfo {volume}
  {36}},\ \bibinfo {pages} {4922} (\bibinfo {year} {1987})}\BibitemShut
  {NoStop}%
\bibitem [{\citenamefont {Chalker}\ and\ \citenamefont
  {Mehlig}(1998)}]{chalker_mehlig_prl1998}%
  \BibitemOpen
  \bibfield  {author} {\bibinfo {author} {\bibfnamefont {J.~T.}\ \bibnamefont
  {Chalker}}\ and\ \bibinfo {author} {\bibfnamefont {B.}~\bibnamefont
  {Mehlig}},\ }\href {\doibase 10.1103/PhysRevLett.81.3367} {\bibfield
  {journal} {\bibinfo  {journal} {Phys. Rev. Lett.}\ }\textbf {\bibinfo
  {volume} {81}},\ \bibinfo {pages} {3367} (\bibinfo {year}
  {1998})}\BibitemShut {NoStop}%
\bibitem [{\citenamefont {Mehlig}\ and\ \citenamefont
  {Chalker}(2000)}]{mehlig_chalker_jmathphys2000}%
  \BibitemOpen
  \bibfield  {author} {\bibinfo {author} {\bibfnamefont {B.}~\bibnamefont
  {Mehlig}}\ and\ \bibinfo {author} {\bibfnamefont {J.~T.}\ \bibnamefont
  {Chalker}},\ }\href {\doibase 10.1063/1.533302} {\bibfield  {journal}
  {\bibinfo  {journal} {J. Math Phys.}\ }\textbf {\bibinfo {volume} {41}},\
  \bibinfo {pages} {3233} (\bibinfo {year} {2000})}\BibitemShut {NoStop}%
\bibitem [{\citenamefont {Huang}\ and\ \citenamefont
  {Doiron}(2017)}]{huang_doiron_coinb2017}%
  \BibitemOpen
  \bibfield  {author} {\bibinfo {author} {\bibfnamefont {C.}~\bibnamefont
  {Huang}}\ and\ \bibinfo {author} {\bibfnamefont {B.}~\bibnamefont {Doiron}},\
  }\href {\doibase https://doi.org/10.1016/j.conb.2017.07.003} {\bibfield
  {journal} {\bibinfo  {journal} {Curr. Opin. Neurobiol.}\ }\textbf {\bibinfo
  {volume} {46}},\ \bibinfo {pages} {31} (\bibinfo {year} {2017})}\BibitemShut
  {NoStop}%
\bibitem [{\citenamefont {Rajan}\ \emph {et~al.}(2010)\citenamefont {Rajan},
  \citenamefont {Abbott},\ and\ \citenamefont
  {Sompolinsky}}]{rajan_abbott_sompolinsky_pre2010}%
  \BibitemOpen
  \bibfield  {author} {\bibinfo {author} {\bibfnamefont {K.}~\bibnamefont
  {Rajan}}, \bibinfo {author} {\bibfnamefont {L.~F.}\ \bibnamefont {Abbott}}, \
  and\ \bibinfo {author} {\bibfnamefont {H.}~\bibnamefont {Sompolinsky}},\
  }\href {\doibase 10.1103/PhysRevE.82.011903} {\bibfield  {journal} {\bibinfo
  {journal} {Phys. Rev. E}\ }\textbf {\bibinfo {volume} {82}},\ \bibinfo
  {pages} {011903} (\bibinfo {year} {2010})}\BibitemShut {NoStop}%
\bibitem [{\citenamefont {Kadmon}\ and\ \citenamefont
  {Sompolinsky}(2015)}]{kadmon_sompolinsky_prx2015}%
  \BibitemOpen
  \bibfield  {author} {\bibinfo {author} {\bibfnamefont {J.}~\bibnamefont
  {Kadmon}}\ and\ \bibinfo {author} {\bibfnamefont {H.}~\bibnamefont
  {Sompolinsky}},\ }\href {\doibase 10.1103/PhysRevX.5.041030} {\bibfield
  {journal} {\bibinfo  {journal} {Phys. Rev. X}\ }\textbf {\bibinfo {volume}
  {5}},\ \bibinfo {pages} {041030} (\bibinfo {year} {2015})}\BibitemShut
  {NoStop}%
\bibitem [{\citenamefont {Harish}\ and\ \citenamefont
  {Hansel}(2015)}]{harish_hansel_ploscb2015}%
  \BibitemOpen
  \bibfield  {author} {\bibinfo {author} {\bibfnamefont {O.}~\bibnamefont
  {Harish}}\ and\ \bibinfo {author} {\bibfnamefont {D.}~\bibnamefont
  {Hansel}},\ }\href {\doibase 10.1371/journal.pcbi.1004266} {\bibfield
  {journal} {\bibinfo  {journal} {{PLoS} Comput. Biol.}\ }\textbf {\bibinfo
  {volume} {11}},\ \bibinfo {pages} {e1004266} (\bibinfo {year}
  {2015})}\BibitemShut {NoStop}%
\bibitem [{\citenamefont {Mastrogiuseppe}\ and\ \citenamefont
  {Ostojic}(2017)}]{mastrogiuseppe_ostojic_2016}%
  \BibitemOpen
  \bibfield  {author} {\bibinfo {author} {\bibfnamefont {F.}~\bibnamefont
  {Mastrogiuseppe}}\ and\ \bibinfo {author} {\bibfnamefont {S.}~\bibnamefont
  {Ostojic}},\ }\href {\doibase 10.1371/journal.pcbi.1005498} {\bibfield
  {journal} {\bibinfo  {journal} {PLOS Comput. Biol.}\ }\textbf {\bibinfo
  {volume} {13}},\ \bibinfo {pages} {1} (\bibinfo {year} {2017})}\BibitemShut
  {NoStop}%
\bibitem [{\citenamefont {Wainrib}\ and\ \citenamefont
  {Touboul}(2013)}]{wainrib_touboul_prl2013}%
  \BibitemOpen
  \bibfield  {author} {\bibinfo {author} {\bibfnamefont {G.}~\bibnamefont
  {Wainrib}}\ and\ \bibinfo {author} {\bibfnamefont {J.}~\bibnamefont
  {Touboul}},\ }\href@noop {} {\bibfield  {journal} {\bibinfo  {journal} {Phys.
  Rev. Lett.}\ }\textbf {\bibinfo {volume} {110}},\ \bibinfo {pages} {118101}
  (\bibinfo {year} {2013})}\BibitemShut {NoStop}%
\bibitem [{\citenamefont {Ginibre}(1965)}]{ginibre_jmatphys1965}%
  \BibitemOpen
  \bibfield  {author} {\bibinfo {author} {\bibfnamefont {J.}~\bibnamefont
  {Ginibre}},\ }\href@noop {} {\bibfield  {journal} {\bibinfo  {journal} {J.
  Math Phys.}\ }\textbf {\bibinfo {volume} {6}},\ \bibinfo {pages} {440}
  (\bibinfo {year} {1965})}\BibitemShut {NoStop}%
\bibitem [{\citenamefont {Girko}(1984)}]{girko_thappprob1984}%
  \BibitemOpen
  \bibfield  {author} {\bibinfo {author} {\bibfnamefont {V.~L.}\ \bibnamefont
  {Girko}},\ }\href@noop {} {\bibfield  {journal} {\bibinfo  {journal} {Teor.
  Veroyatnost. i Primenen.}\ }\textbf {\bibinfo {volume} {29}},\ \bibinfo
  {pages} {669} (\bibinfo {year} {1984})}\BibitemShut {NoStop}%
\bibitem [{\citenamefont {Tao}\ \emph {et~al.}(2010)\citenamefont {Tao},
  \citenamefont {Vu},\ and\ \citenamefont {Krishnapur}}]{tao_vu_annprob2010}%
  \BibitemOpen
  \bibfield  {author} {\bibinfo {author} {\bibfnamefont {T.}~\bibnamefont
  {Tao}}, \bibinfo {author} {\bibfnamefont {V.}~\bibnamefont {Vu}}, \ and\
  \bibinfo {author} {\bibfnamefont {M.}~\bibnamefont {Krishnapur}},\ }\href
  {\doibase 10.1214/10-AOP534} {\bibfield  {journal} {\bibinfo  {journal} {Ann.
  Prob.}\ }\textbf {\bibinfo {volume} {38}},\ \bibinfo {pages} {2023} (\bibinfo
  {year} {2010})}\BibitemShut {NoStop}%
\bibitem [{\citenamefont {Girko}(1985)}]{girko_teorver1985}%
  \BibitemOpen
  \bibfield  {author} {\bibinfo {author} {\bibfnamefont {V.~L.}\ \bibnamefont
  {Girko}},\ }\href@noop {} {\bibfield  {journal} {\bibinfo  {journal} {Teor.
  Veroyatnost. i Primenen.}\ }\textbf {\bibinfo {volume} {30}},\ \bibinfo
  {pages} {640} (\bibinfo {year} {1985})}\BibitemShut {NoStop}%
\bibitem [{\citenamefont {Sommers}\ \emph {et~al.}(1988)\citenamefont
  {Sommers}, \citenamefont {Crisanti}, \citenamefont {Sompolinsky},\ and\
  \citenamefont {Stein}}]{sommers_etal_prl1988}%
  \BibitemOpen
  \bibfield  {author} {\bibinfo {author} {\bibfnamefont {H.~J.}\ \bibnamefont
  {Sommers}}, \bibinfo {author} {\bibfnamefont {A.}~\bibnamefont {Crisanti}},
  \bibinfo {author} {\bibfnamefont {H.}~\bibnamefont {Sompolinsky}}, \ and\
  \bibinfo {author} {\bibfnamefont {Y.}~\bibnamefont {Stein}},\ }\href
  {\doibase 10.1103/PhysRevLett.61.259} {\bibfield  {journal} {\bibinfo
  {journal} {Phys. Rev. Lett.}\ }\textbf {\bibinfo {volume} {60}},\ \bibinfo
  {pages} {1895} (\bibinfo {year} {1988})}\BibitemShut {NoStop}%
\bibitem [{\citenamefont {{Naumov}}(2013)}]{naumov_arxiv2012}%
  \BibitemOpen
  \bibfield  {author} {\bibinfo {author} {\bibfnamefont {A.}~\bibnamefont
  {{Naumov}}},\ }\href@noop {} {\bibfield  {journal} {\bibinfo  {journal}
  {Vestnik Moskov. Univ. Ser. XV Vychisl. Mat. Kibernet.}\ }\textbf {\bibinfo
  {volume} {1}},\ \bibinfo {pages} {31} (\bibinfo {year} {2013})}\BibitemShut
  {NoStop}%
\bibitem [{\citenamefont {Nguyen}\ and\ \citenamefont
  {O'Rourke}(2015)}]{nguyen_orourke_imrn2014}%
  \BibitemOpen
  \bibfield  {author} {\bibinfo {author} {\bibfnamefont {H.~H.}\ \bibnamefont
  {Nguyen}}\ and\ \bibinfo {author} {\bibfnamefont {S.}~\bibnamefont
  {O'Rourke}},\ }\href@noop {} {\bibfield  {journal} {\bibinfo  {journal} {Int.
  Math. Res. Not.}\ ,\ \bibinfo {pages} {7620}} (\bibinfo {year}
  {2015})}\BibitemShut {NoStop}%
\bibitem [{\citenamefont {{Schuecker}}\ \emph {et~al.}(2016)\citenamefont
  {{Schuecker}}, \citenamefont {{Goedeke}},\ and\ \citenamefont
  {{Helias}}}]{schuecker_goedeke_helias_arxiv2016}%
  \BibitemOpen
  \bibfield  {author} {\bibinfo {author} {\bibfnamefont {J.}~\bibnamefont
  {{Schuecker}}}, \bibinfo {author} {\bibfnamefont {S.}~\bibnamefont
  {{Goedeke}}}, \ and\ \bibinfo {author} {\bibfnamefont {M.}~\bibnamefont
  {{Helias}}},\ }\href@noop {} {\bibfield  {journal} {\bibinfo  {journal}
  {ArXiv e-prints}\ } (\bibinfo {year} {2016})},\ \Eprint
  {http://arxiv.org/abs/1603.01880} {arXiv:1603.01880 [q-bio.NC]} \BibitemShut
  {NoStop}%
\bibitem [{\citenamefont {Trefethen}\ and\ \citenamefont
  {Embree}(2005)}]{trefethen_embree_2005}%
  \BibitemOpen
  \bibfield  {author} {\bibinfo {author} {\bibfnamefont {L.~N.}\ \bibnamefont
  {Trefethen}}\ and\ \bibinfo {author} {\bibfnamefont {M.}~\bibnamefont
  {Embree}},\ }\href@noop {} {\emph {\bibinfo {title} {Spectra and
  pseudospectra : the behavior of nonnormal matrices and operators}}}\
  (\bibinfo  {publisher} {Princeton University Press},\ \bibinfo {address}
  {Princeton, Oxford},\ \bibinfo {year} {2005})\BibitemShut {NoStop}%
\bibitem [{\citenamefont {Sompolinsky}\ and\ \citenamefont
  {Zippelius}(1982)}]{sompolinsky_zippelius_prb1982}%
  \BibitemOpen
  \bibfield  {author} {\bibinfo {author} {\bibfnamefont {H.}~\bibnamefont
  {Sompolinsky}}\ and\ \bibinfo {author} {\bibfnamefont {A.}~\bibnamefont
  {Zippelius}},\ }\href {\doibase 10.1103/PhysRevB.25.6860} {\bibfield
  {journal} {\bibinfo  {journal} {Phys. Rev. B}\ }\textbf {\bibinfo {volume}
  {25}},\ \bibinfo {pages} {6860} (\bibinfo {year} {1982})}\BibitemShut
  {NoStop}%
\bibitem [{\citenamefont {Bravi}\ \emph {et~al.}(2016)\citenamefont {Bravi},
  \citenamefont {Sollich},\ and\ \citenamefont
  {Opper}}]{bravi_sollich_opper_jpa2016}%
  \BibitemOpen
  \bibfield  {author} {\bibinfo {author} {\bibfnamefont {B.}~\bibnamefont
  {Bravi}}, \bibinfo {author} {\bibfnamefont {P.}~\bibnamefont {Sollich}}, \
  and\ \bibinfo {author} {\bibfnamefont {M.}~\bibnamefont {Opper}},\ }\href
  {http://stacks.iop.org/1751-8121/49/i=19/a=194003} {\bibfield  {journal}
  {\bibinfo  {journal} {J. Phys. A-Math.}\ }\textbf {\bibinfo {volume} {49}},\
  \bibinfo {pages} {194003} (\bibinfo {year} {2016})}\BibitemShut {NoStop}%
\bibitem [{\citenamefont {Hennequin}\ \emph {et~al.}(2012)\citenamefont
  {Hennequin}, \citenamefont {Vogels},\ and\ \citenamefont
  {Gerstner}}]{hennequin_vogels_gerstner_pre2012}%
  \BibitemOpen
  \bibfield  {author} {\bibinfo {author} {\bibfnamefont {G.}~\bibnamefont
  {Hennequin}}, \bibinfo {author} {\bibfnamefont {T.~P.}\ \bibnamefont
  {Vogels}}, \ and\ \bibinfo {author} {\bibfnamefont {W.}~\bibnamefont
  {Gerstner}},\ }\href {\doibase 10.1103/PhysRevE.86.011909} {\bibfield
  {journal} {\bibinfo  {journal} {Phys. Rev. E}\ }\textbf {\bibinfo {volume}
  {86}},\ \bibinfo {pages} {011909} (\bibinfo {year} {2012})}\BibitemShut
  {NoStop}%
\bibitem [{\citenamefont {Bouchaud}\ \emph {et~al.}(1998)\citenamefont
  {Bouchaud}, \citenamefont {Cugliandolo}, \citenamefont {Kurchan},\ and\
  \citenamefont {M{\'e}zard}}]{bouchaud_etal_sprf1998}%
  \BibitemOpen
  \bibfield  {author} {\bibinfo {author} {\bibfnamefont {J.-P.}\ \bibnamefont
  {Bouchaud}}, \bibinfo {author} {\bibfnamefont {L.~F.}\ \bibnamefont
  {Cugliandolo}}, \bibinfo {author} {\bibfnamefont {J.}~\bibnamefont
  {Kurchan}}, \ and\ \bibinfo {author} {\bibfnamefont {M.}~\bibnamefont
  {M{\'e}zard}},\ }in\ \href@noop {} {\emph {\bibinfo {booktitle} {Spin glasses
  and random fields}}},\ \bibinfo {editor} {edited by\ \bibinfo {editor}
  {\bibfnamefont {A.~P.}\ \bibnamefont {Young}}}\ (\bibinfo  {publisher} {World
  Scientific, Singapore},\ \bibinfo {year} {1998})\ Chap.~\bibinfo {chapter}
  {6}, pp.\ \bibinfo {pages} {161--223}\BibitemShut {NoStop}%
\bibitem [{\citenamefont {Hopfield}(1982)}]{hopfield_pnas1982}%
  \BibitemOpen
  \bibfield  {author} {\bibinfo {author} {\bibfnamefont {J.~J.}\ \bibnamefont
  {Hopfield}},\ }\href {\doibase 10.1073/pnas.79.8.2554} {\bibfield  {journal}
  {\bibinfo  {journal} {P. Natl. Acad. Sci. USA}\ }\textbf {\bibinfo {volume}
  {79}},\ \bibinfo {pages} {2554} (\bibinfo {year} {1982})}\BibitemShut
  {NoStop}%
\bibitem [{\citenamefont {Amit}\ \emph {et~al.}(1985)\citenamefont {Amit},
  \citenamefont {Gutfreund},\ and\ \citenamefont
  {Sompolinsky}}]{amit_etal_pra1985}%
  \BibitemOpen
  \bibfield  {author} {\bibinfo {author} {\bibfnamefont {D.~J.}\ \bibnamefont
  {Amit}}, \bibinfo {author} {\bibfnamefont {H.}~\bibnamefont {Gutfreund}}, \
  and\ \bibinfo {author} {\bibfnamefont {H.}~\bibnamefont {Sompolinsky}},\
  }\href {\doibase 10.1103/physreva.32.1007} {\bibfield  {journal} {\bibinfo
  {journal} {Phys. Rev. A}\ }\textbf {\bibinfo {volume} {32}},\ \bibinfo
  {pages} {1007} (\bibinfo {year} {1985})}\BibitemShut {NoStop}%
\bibitem [{\citenamefont {Hertz}\ \emph {et~al.}(1986)\citenamefont {Hertz},
  \citenamefont {Grinstein},\ and\ \citenamefont
  {Solla}}]{hertz_grinstein_solla_aip1986}%
  \BibitemOpen
  \bibfield  {author} {\bibinfo {author} {\bibfnamefont {J.~A.}\ \bibnamefont
  {Hertz}}, \bibinfo {author} {\bibfnamefont {G.}~\bibnamefont {Grinstein}}, \
  and\ \bibinfo {author} {\bibfnamefont {S.~A.}\ \bibnamefont {Solla}},\ }in\
  \href {\doibase 10.1063/1.36259} {\emph {\bibinfo {booktitle} {American
  Institute of Physics Conference Proceedings on Neural Networks for
  Computing}}},\ Vol.\ \bibinfo {volume} {151},\ \bibinfo {editor} {edited by\
  \bibinfo {editor} {\bibfnamefont {J.~S.}\ \bibnamefont {Denker}}}\ (\bibinfo
  {publisher} {American Institute of Physics Inc.},\ \bibinfo {year} {1986})\
  pp.\ \bibinfo {pages} {212--218}\BibitemShut {NoStop}%
\bibitem [{\citenamefont {Derrida}\ \emph {et~al.}(1987)\citenamefont
  {Derrida}, \citenamefont {Gardner},\ and\ \citenamefont
  {Zippelius}}]{derrida_gardner_zippelius_europhyslett1987}%
  \BibitemOpen
  \bibfield  {author} {\bibinfo {author} {\bibfnamefont {B.}~\bibnamefont
  {Derrida}}, \bibinfo {author} {\bibfnamefont {E.}~\bibnamefont {Gardner}}, \
  and\ \bibinfo {author} {\bibfnamefont {A.}~\bibnamefont {Zippelius}},\ }\href
  {\doibase 10.1209/0295-5075/4/2/007} {\bibfield  {journal} {\bibinfo
  {journal} {Europhys. Lett.}\ }\textbf {\bibinfo {volume} {4}},\ \bibinfo
  {pages} {167} (\bibinfo {year} {1987})}\BibitemShut {NoStop}%
\bibitem [{\citenamefont {van Vreeswijk}\ and\ \citenamefont
  {Sompolinsky}(1996)}]{vanvreeswijk_sompolinsky_science1996}%
  \BibitemOpen
  \bibfield  {author} {\bibinfo {author} {\bibfnamefont {C.}~\bibnamefont {van
  Vreeswijk}}\ and\ \bibinfo {author} {\bibfnamefont {H.}~\bibnamefont
  {Sompolinsky}},\ }\href {\doibase 10.1126/science.274.5293.1724} {\bibfield
  {journal} {\bibinfo  {journal} {Science}\ }\textbf {\bibinfo {volume}
  {274}},\ \bibinfo {pages} {1724} (\bibinfo {year} {1996})}\BibitemShut
  {NoStop}%
\bibitem [{\citenamefont {Litwin-Kumar}\ and\ \citenamefont
  {Doiron}(2012)}]{litwin-kumar_doiron_natns2012}%
  \BibitemOpen
  \bibfield  {author} {\bibinfo {author} {\bibfnamefont {A.}~\bibnamefont
  {Litwin-Kumar}}\ and\ \bibinfo {author} {\bibfnamefont {B.}~\bibnamefont
  {Doiron}},\ }\href {\doibase 10.1038/nn.3220} {\bibfield  {journal} {\bibinfo
   {journal} {Nat. Neurosci.}\ }\textbf {\bibinfo {volume} {15}},\ \bibinfo
  {pages} {1498} (\bibinfo {year} {2012})}\BibitemShut {NoStop}%
\bibitem [{\citenamefont {Stern}\ \emph {et~al.}(2014)\citenamefont {Stern},
  \citenamefont {Sompolinsky},\ and\ \citenamefont
  {Abbott}}]{stern_sompolinsky_abbott_pre2014}%
  \BibitemOpen
  \bibfield  {author} {\bibinfo {author} {\bibfnamefont {M.}~\bibnamefont
  {Stern}}, \bibinfo {author} {\bibfnamefont {H.}~\bibnamefont {Sompolinsky}},
  \ and\ \bibinfo {author} {\bibfnamefont {L.~F.}\ \bibnamefont {Abbott}},\
  }\href {\doibase 10.1103/PhysRevE.90.062710} {\bibfield  {journal} {\bibinfo
  {journal} {Phys. Rev. E}\ }\textbf {\bibinfo {volume} {90}},\ \bibinfo
  {pages} {062710} (\bibinfo {year} {2014})}\BibitemShut {NoStop}%
\bibitem [{\citenamefont {Roxin}(2011)}]{roxin_ficn2011}%
  \BibitemOpen
  \bibfield  {author} {\bibinfo {author} {\bibfnamefont {A.}~\bibnamefont
  {Roxin}},\ }\href {\doibase 10.3389/fncom.2011.00008} {\bibfield  {journal}
  {\bibinfo  {journal} {Front. Comput. Neurosci.}\ }\textbf {\bibinfo {volume}
  {5}},\ \bibinfo {pages} {8} (\bibinfo {year} {2011})}\BibitemShut {NoStop}%
\bibitem [{\citenamefont {Zhao}\ \emph {et~al.}(2011)\citenamefont {Zhao},
  \citenamefont {Beverlin}, \citenamefont {Netoff},\ and\ \citenamefont
  {Nykamp}}]{zhao_etal_ficns2011}%
  \BibitemOpen
  \bibfield  {author} {\bibinfo {author} {\bibfnamefont {L.}~\bibnamefont
  {Zhao}}, \bibinfo {author} {\bibfnamefont {B.}~\bibnamefont {Beverlin}},
  \bibinfo {author} {\bibfnamefont {T.}~\bibnamefont {Netoff}}, \ and\ \bibinfo
  {author} {\bibfnamefont {D.}~\bibnamefont {Nykamp}},\ }\href {\doibase
  10.3389/fncom.2011.00028} {\bibfield  {journal} {\bibinfo  {journal} {Front.
  Comput. Neurosci.}\ }\textbf {\bibinfo {volume} {5}},\ \bibinfo {pages} {28}
  (\bibinfo {year} {2011})}\BibitemShut {NoStop}%
\bibitem [{\citenamefont {Bimbard}\ \emph {et~al.}(2016)\citenamefont
  {Bimbard}, \citenamefont {Ledoux},\ and\ \citenamefont
  {Ostojic}}]{bimbard_ledoux_ostojic_pre2016}%
  \BibitemOpen
  \bibfield  {author} {\bibinfo {author} {\bibfnamefont {C.}~\bibnamefont
  {Bimbard}}, \bibinfo {author} {\bibfnamefont {E.}~\bibnamefont {Ledoux}}, \
  and\ \bibinfo {author} {\bibfnamefont {S.}~\bibnamefont {Ostojic}},\
  }\href@noop {} {\bibfield  {journal} {\bibinfo  {journal} {Phys. Rev. E}\
  }\textbf {\bibinfo {volume} {94}},\ \bibinfo {pages} {062207} (\bibinfo
  {year} {2016})}\BibitemShut {NoStop}%
\bibitem [{\citenamefont {Brunel}(2016)}]{brunel_natns2016}%
  \BibitemOpen
  \bibfield  {author} {\bibinfo {author} {\bibfnamefont {N.}~\bibnamefont
  {Brunel}},\ }\href {\doibase 10.1038/nn.4286} {\bibfield  {journal} {\bibinfo
   {journal} {Nat. Neurosci.}\ }\textbf {\bibinfo {volume} {19}},\ \bibinfo
  {pages} {749} (\bibinfo {year} {2016})}\BibitemShut {NoStop}%
\bibitem [{\citenamefont {Ganguli}\ \emph {et~al.}(2008)\citenamefont
  {Ganguli}, \citenamefont {Huh},\ and\ \citenamefont
  {Sompolinsky}}]{ganguli_sompolinsky}%
  \BibitemOpen
  \bibfield  {author} {\bibinfo {author} {\bibfnamefont {S.}~\bibnamefont
  {Ganguli}}, \bibinfo {author} {\bibfnamefont {D.}~\bibnamefont {Huh}}, \ and\
  \bibinfo {author} {\bibfnamefont {H.}~\bibnamefont {Sompolinsky}},\ }\href
  {\doibase 10.1073/pnas.0804451105} {\bibfield  {journal} {\bibinfo  {journal}
  {P. Natl. Acad. Sci. USA}\ }\textbf {\bibinfo {volume} {105}},\ \bibinfo
  {pages} {18970} (\bibinfo {year} {2008})}\BibitemShut {NoStop}%
\bibitem [{\citenamefont {Murphy}\ and\ \citenamefont
  {Miller}(2009)}]{murphy_miller_neuron2009}%
  \BibitemOpen
  \bibfield  {author} {\bibinfo {author} {\bibfnamefont {B.~K.}\ \bibnamefont
  {Murphy}}\ and\ \bibinfo {author} {\bibfnamefont {K.~D.}\ \bibnamefont
  {Miller}},\ }\href {\doibase 10.1016/j.neuron.2009.02.005} {\bibfield
  {journal} {\bibinfo  {journal} {Neuron}\ }\textbf {\bibinfo {volume} {61}},\
  \bibinfo {pages} {635} (\bibinfo {year} {2009})}\BibitemShut {NoStop}%
\bibitem [{\citenamefont {Goldman}(2008)}]{goldman_neuron2008}%
  \BibitemOpen
  \bibfield  {author} {\bibinfo {author} {\bibfnamefont {M.~S.}\ \bibnamefont
  {Goldman}},\ }\href {\doibase 10.1016/j.neuron.2008.12.012} {\bibfield
  {journal} {\bibinfo  {journal} {Neuron}\ }\textbf {\bibinfo {volume} {61}},\
  \bibinfo {pages} {621} (\bibinfo {year} {2008})}\BibitemShut {NoStop}%
\bibitem [{\citenamefont {Hennequin}\ \emph {et~al.}(2014)\citenamefont
  {Hennequin}, \citenamefont {Vogels},\ and\ \citenamefont
  {Gerstner}}]{hennequin_vogels_gerstner_neuron2014}%
  \BibitemOpen
  \bibfield  {author} {\bibinfo {author} {\bibfnamefont {G.}~\bibnamefont
  {Hennequin}}, \bibinfo {author} {\bibfnamefont {T.~P.}\ \bibnamefont
  {Vogels}}, \ and\ \bibinfo {author} {\bibfnamefont {W.}~\bibnamefont
  {Gerstner}},\ }\href@noop {} {\bibfield  {journal} {\bibinfo  {journal}
  {Neuron}\ }\textbf {\bibinfo {volume} {82}},\ \bibinfo {pages} {1394}
  (\bibinfo {year} {2014})}\BibitemShut {NoStop}%
\bibitem [{\citenamefont {Ahmadian}\ \emph {et~al.}(2015)\citenamefont
  {Ahmadian}, \citenamefont {Fumarola},\ and\ \citenamefont
  {Miller}}]{ahmadian_etal_pre2015}%
  \BibitemOpen
  \bibfield  {author} {\bibinfo {author} {\bibfnamefont {Y.}~\bibnamefont
  {Ahmadian}}, \bibinfo {author} {\bibfnamefont {F.}~\bibnamefont {Fumarola}},
  \ and\ \bibinfo {author} {\bibfnamefont {K.~D.}\ \bibnamefont {Miller}},\
  }\href {\doibase 10.1103/PhysRevE.91.012820} {\bibfield  {journal} {\bibinfo
  {journal} {Phys. Rev. E}\ }\textbf {\bibinfo {volume} {91}},\ \bibinfo
  {pages} {012820} (\bibinfo {year} {2015})}\BibitemShut {NoStop}%
\bibitem [{\citenamefont {M{\'e}zard}\ \emph {et~al.}(1987)\citenamefont
  {M{\'e}zard}, \citenamefont {Parisi},\ and\ \citenamefont
  {Virasoro}}]{mezard_parisi_virasoro_spinglasses1987}%
  \BibitemOpen
  \bibfield  {author} {\bibinfo {author} {\bibfnamefont {M.}~\bibnamefont
  {M{\'e}zard}}, \bibinfo {author} {\bibfnamefont {G.}~\bibnamefont {Parisi}},
  \ and\ \bibinfo {author} {\bibfnamefont {M.}~\bibnamefont {Virasoro}},\
  }\href@noop {} {\emph {\bibinfo {title} {Spin glass theory and beyond: An
  Introduction to the Replica Method and Its Applications}}},\ Vol.~\bibinfo
  {volume} {9}\ (\bibinfo  {publisher} {World Scientific},\ \bibinfo {year}
  {1987})\BibitemShut {NoStop}%
\bibitem [{\citenamefont {Cugliandolo}\ and\ \citenamefont
  {Kurchan}(1993)}]{cugliandolo_kurchan_prl1993}%
  \BibitemOpen
  \bibfield  {author} {\bibinfo {author} {\bibfnamefont {L.~F.}\ \bibnamefont
  {Cugliandolo}}\ and\ \bibinfo {author} {\bibfnamefont {J.}~\bibnamefont
  {Kurchan}},\ }\href {\doibase 10.1103/PhysRevLett.71.173} {\bibfield
  {journal} {\bibinfo  {journal} {Phys. Rev. Lett.}\ }\textbf {\bibinfo
  {volume} {71}},\ \bibinfo {pages} {173} (\bibinfo {year} {1993})}\BibitemShut
  {NoStop}%
\bibitem [{\citenamefont {Cugliandolo}\ and\ \citenamefont
  {Dean}(1995)}]{cugliandolo_dean_pra1995}%
  \BibitemOpen
  \bibfield  {author} {\bibinfo {author} {\bibfnamefont {L.~F.}\ \bibnamefont
  {Cugliandolo}}\ and\ \bibinfo {author} {\bibfnamefont {D.~S.}\ \bibnamefont
  {Dean}},\ }\href {\doibase 10.1088/0305-4470/28/15/003} {\bibfield  {journal}
  {\bibinfo  {journal} {J. Phys. A-Math. Gen.}\ }\textbf {\bibinfo {volume}
  {28}},\ \bibinfo {pages} {4213} (\bibinfo {year} {1995})}\BibitemShut
  {NoStop}%
\bibitem [{\citenamefont {Iori}\ and\ \citenamefont
  {Marinari}(1997)}]{iori_marinari_jpa1997}%
  \BibitemOpen
  \bibfield  {author} {\bibinfo {author} {\bibfnamefont {G.}~\bibnamefont
  {Iori}}\ and\ \bibinfo {author} {\bibfnamefont {E.}~\bibnamefont
  {Marinari}},\ }\href@noop {} {\bibfield  {journal} {\bibinfo  {journal} {J.
  Phys. A-Math. Gen.}\ }\textbf {\bibinfo {volume} {30}},\ \bibinfo {pages}
  {4489} (\bibinfo {year} {1997})}\BibitemShut {NoStop}%
\bibitem [{\citenamefont {Marinari}\ and\ \citenamefont
  {Stariolo}(1998)}]{marinari_stariolo_jpa1998}%
  \BibitemOpen
  \bibfield  {author} {\bibinfo {author} {\bibfnamefont {E.}~\bibnamefont
  {Marinari}}\ and\ \bibinfo {author} {\bibfnamefont {D.~A.}\ \bibnamefont
  {Stariolo}},\ }\href@noop {} {\bibfield  {journal} {\bibinfo  {journal} {J.
  Phys. A-Math. Gen.}\ }\textbf {\bibinfo {volume} {31}},\ \bibinfo {pages}
  {5021} (\bibinfo {year} {1998})}\BibitemShut {NoStop}%
\bibitem [{\citenamefont {Lefort}\ \emph {et~al.}(2009)\citenamefont {Lefort},
  \citenamefont {Tomm}, \citenamefont {Sarria},\ and\ \citenamefont
  {Petersen}}]{lefort_etal_neuron2009}%
  \BibitemOpen
  \bibfield  {author} {\bibinfo {author} {\bibfnamefont {S.}~\bibnamefont
  {Lefort}}, \bibinfo {author} {\bibfnamefont {C.}~\bibnamefont {Tomm}},
  \bibinfo {author} {\bibfnamefont {J.~C.~F.}\ \bibnamefont {Sarria}}, \ and\
  \bibinfo {author} {\bibfnamefont {C.~C.}\ \bibnamefont {Petersen}},\
  }\href@noop {} {\bibfield  {journal} {\bibinfo  {journal} {Neuron}\ }\textbf
  {\bibinfo {volume} {61}},\ \bibinfo {pages} {301} (\bibinfo {year}
  {2009})}\BibitemShut {NoStop}%
\bibitem [{\citenamefont {Murray}\ \emph {et~al.}(2014)\citenamefont {Murray},
  \citenamefont {Bernacchia}, \citenamefont {Freedman}, \citenamefont {Romo},
  \citenamefont {Wallis}, \citenamefont {Cai}, \citenamefont {Padoa-Schioppa},
  \citenamefont {Pasternak}, \citenamefont {Seo}, \citenamefont {Lee},\ and\
  \citenamefont {Wang}}]{murray_etal_natns2014}%
  \BibitemOpen
  \bibfield  {author} {\bibinfo {author} {\bibfnamefont {J.~D.}\ \bibnamefont
  {Murray}}, \bibinfo {author} {\bibfnamefont {A.}~\bibnamefont {Bernacchia}},
  \bibinfo {author} {\bibfnamefont {D.~J.}\ \bibnamefont {Freedman}}, \bibinfo
  {author} {\bibfnamefont {R.}~\bibnamefont {Romo}}, \bibinfo {author}
  {\bibfnamefont {J.~D.}\ \bibnamefont {Wallis}}, \bibinfo {author}
  {\bibfnamefont {X.}~\bibnamefont {Cai}}, \bibinfo {author} {\bibfnamefont
  {C.}~\bibnamefont {Padoa-Schioppa}}, \bibinfo {author} {\bibfnamefont
  {T.}~\bibnamefont {Pasternak}}, \bibinfo {author} {\bibfnamefont
  {H.}~\bibnamefont {Seo}}, \bibinfo {author} {\bibfnamefont {D.}~\bibnamefont
  {Lee}}, \ and\ \bibinfo {author} {\bibfnamefont {X.~J.}\ \bibnamefont
  {Wang}},\ }\href@noop {} {\bibfield  {journal} {\bibinfo  {journal} {Nat.
  Neurosci.}\ }\textbf {\bibinfo {volume} {17}},\ \bibinfo {pages} {1661}
  (\bibinfo {year} {2014})}\BibitemShut {NoStop}%
\bibitem [{\citenamefont {Wang}\ \emph {et~al.}(2008)\citenamefont {Wang},
  \citenamefont {Stradtman}, \citenamefont {Wang},\ and\ \citenamefont
  {Gao}}]{wang_etal_pnas08}%
  \BibitemOpen
  \bibfield  {author} {\bibinfo {author} {\bibfnamefont {H.}~\bibnamefont
  {Wang}}, \bibinfo {author} {\bibfnamefont {G.~G.}\ \bibnamefont {Stradtman}},
  \bibinfo {author} {\bibfnamefont {X.~J.}\ \bibnamefont {Wang}}, \ and\
  \bibinfo {author} {\bibfnamefont {W.~J.}\ \bibnamefont {Gao}},\ }\href@noop
  {} {\bibfield  {journal} {\bibinfo  {journal} {P. Natl. Acad. Sci. USA}\
  }\textbf {\bibinfo {volume} {105}},\ \bibinfo {pages} {16791} (\bibinfo
  {year} {2008})}\BibitemShut {NoStop}%
\bibitem [{\citenamefont {Elston}(2003)}]{elston_cercor2003}%
  \BibitemOpen
  \bibfield  {author} {\bibinfo {author} {\bibfnamefont {G.~N.}\ \bibnamefont
  {Elston}},\ }\href {\doibase 10.1093/cercor/bhg093} {\bibfield  {journal}
  {\bibinfo  {journal} {Cereb. Cortex}\ }\textbf {\bibinfo {volume} {13}},\
  \bibinfo {pages} {1124} (\bibinfo {year} {2003})}\BibitemShut {NoStop}%
\bibitem [{\citenamefont {Rajan}\ and\ \citenamefont
  {Abbott}(2006)}]{rajan_abbott_prl2006}%
  \BibitemOpen
  \bibfield  {author} {\bibinfo {author} {\bibfnamefont {K.}~\bibnamefont
  {Rajan}}\ and\ \bibinfo {author} {\bibfnamefont {L.~F.}\ \bibnamefont
  {Abbott}},\ }\href {\doibase 10.1103/PhysRevLett.97.188104} {\bibfield
  {journal} {\bibinfo  {journal} {Phys. Rev. Lett.}\ }\textbf {\bibinfo
  {volume} {97}},\ \bibinfo {pages} {188104} (\bibinfo {year}
  {2006})}\BibitemShut {NoStop}%
\bibitem [{\citenamefont {Ostojic}(2014)}]{ostojic_natns2014}%
  \BibitemOpen
  \bibfield  {author} {\bibinfo {author} {\bibfnamefont {S.}~\bibnamefont
  {Ostojic}},\ }\href@noop {} {\bibfield  {journal} {\bibinfo  {journal} {Nat.
  Neurosci.}\ }\textbf {\bibinfo {volume} {17}},\ \bibinfo {pages} {594}
  (\bibinfo {year} {2014})}\BibitemShut {NoStop}%
\bibitem [{\citenamefont {Aljadeff}\ \emph
  {et~al.}(2015{\natexlab{a}})\citenamefont {Aljadeff}, \citenamefont {Stern},\
  and\ \citenamefont {Sharpee}}]{aljadeff_stern_sharpee_prl2015}%
  \BibitemOpen
  \bibfield  {author} {\bibinfo {author} {\bibfnamefont {J.}~\bibnamefont
  {Aljadeff}}, \bibinfo {author} {\bibfnamefont {M.}~\bibnamefont {Stern}}, \
  and\ \bibinfo {author} {\bibfnamefont {T.}~\bibnamefont {Sharpee}},\ }\href
  {\doibase 10.1103/PhysRevLett.114.088101} {\bibfield  {journal} {\bibinfo
  {journal} {Phys. Rev. Lett.}\ }\textbf {\bibinfo {volume} {114}},\ \bibinfo
  {pages} {088101} (\bibinfo {year} {2015}{\natexlab{a}})}\BibitemShut
  {NoStop}%
\bibitem [{\citenamefont {Kuczala}\ and\ \citenamefont
  {Sharpee}(2016)}]{kuczala_sharpee_pre2016}%
  \BibitemOpen
  \bibfield  {author} {\bibinfo {author} {\bibfnamefont {A.}~\bibnamefont
  {Kuczala}}\ and\ \bibinfo {author} {\bibfnamefont {T.~O.}\ \bibnamefont
  {Sharpee}},\ }\href {\doibase 10.1103/PhysRevE.94.050101} {\bibfield
  {journal} {\bibinfo  {journal} {Phys. Rev. E}\ }\textbf {\bibinfo {volume}
  {94}},\ \bibinfo {pages} {050101} (\bibinfo {year} {2016})}\BibitemShut
  {NoStop}%
\bibitem [{\citenamefont {Aljadeff}\ \emph
  {et~al.}(2015{\natexlab{b}})\citenamefont {Aljadeff}, \citenamefont
  {Renfrew},\ and\ \citenamefont {Stern}}]{aljadeff_renfrew_stern_jmp2015}%
  \BibitemOpen
  \bibfield  {author} {\bibinfo {author} {\bibfnamefont {J.}~\bibnamefont
  {Aljadeff}}, \bibinfo {author} {\bibfnamefont {D.}~\bibnamefont {Renfrew}}, \
  and\ \bibinfo {author} {\bibfnamefont {M.}~\bibnamefont {Stern}},\
  }\href@noop {} {\bibfield  {journal} {\bibinfo  {journal} {J. Math Phys.}\
  }\textbf {\bibinfo {volume} {56}},\ \bibinfo {pages} {103502} (\bibinfo
  {year} {2015}{\natexlab{b}})}\BibitemShut {NoStop}%
\bibitem [{\citenamefont {Prudnikov}\ \emph {et~al.}(1992)\citenamefont
  {Prudnikov}, \citenamefont {Brychkov},\ and\ \citenamefont
  {Marichev}}]{prudnikov_etal}%
  \BibitemOpen
  \bibfield  {author} {\bibinfo {author} {\bibfnamefont {A.~P.}\ \bibnamefont
  {Prudnikov}}, \bibinfo {author} {\bibfnamefont {I.}~\bibnamefont {Brychkov}},
  \ and\ \bibinfo {author} {\bibfnamefont {O.~I.}\ \bibnamefont {Marichev}},\
  }\href@noop {} {\emph {\bibinfo {title} {Integrals and Series [Vol 2 -
  Special Functions]}}}\ (\bibinfo  {publisher} {Gordon and Breach Science
  Publishers},\ \bibinfo {year} {1992})\BibitemShut {NoStop}%
\bibitem [{\citenamefont {Bender}\ and\ \citenamefont
  {Orszag}(1999)}]{bender_orszag_1999}%
  \BibitemOpen
  \bibfield  {author} {\bibinfo {author} {\bibfnamefont {C.~M.}\ \bibnamefont
  {Bender}}\ and\ \bibinfo {author} {\bibfnamefont {S.~A.}\ \bibnamefont
  {Orszag}},\ }\href@noop {} {\emph {\bibinfo {title} {Advanced Mathematical
  Methods for Scientists and Engineers}}}\ (\bibinfo  {publisher} {Springer
  Verlag},\ \bibinfo {year} {1999})\BibitemShut {NoStop}%
\bibitem [{\citenamefont {Gradshteyn}\ and\ \citenamefont
  {Ryzhik}(2007)}]{gradshteyn_ryzhik_2007}%
  \BibitemOpen
  \bibfield  {author} {\bibinfo {author} {\bibfnamefont {I.~S.}\ \bibnamefont
  {Gradshteyn}}\ and\ \bibinfo {author} {\bibfnamefont {I.~M.}\ \bibnamefont
  {Ryzhik}},\ }\href@noop {} {\emph {\bibinfo {title} {Table of integrals,
  series, and products}}},\ \bibinfo {edition} {seventh}\ ed.\ (\bibinfo
  {publisher} {Elsevier/Academic Press, Amsterdam},\ \bibinfo {year} {2007})\
  pp.\ \bibinfo {pages} {xlviii+1171},\ \bibinfo {note} {translated from the
  Russian, Translation edited and with a preface by Alan Jeffrey and Daniel
  Zwillinger, With one CD-ROM (Windows, Macintosh and UNIX)}\BibitemShut
  {NoStop}%
\bibitem [{\citenamefont {Martin}\ \emph {et~al.}(1973)\citenamefont {Martin},
  \citenamefont {Siggia},\ and\ \citenamefont
  {Rose}}]{martin_siggia_rose_pra1973}%
  \BibitemOpen
  \bibfield  {author} {\bibinfo {author} {\bibfnamefont {P.~C.}\ \bibnamefont
  {Martin}}, \bibinfo {author} {\bibfnamefont {E.}~\bibnamefont {Siggia}}, \
  and\ \bibinfo {author} {\bibfnamefont {H.}~\bibnamefont {Rose}},\ }\href
  {\doibase 10.1103/PhysRevA.8.423} {\bibfield  {journal} {\bibinfo  {journal}
  {Phys. Rev. A}\ }\textbf {\bibinfo {volume} {8}},\ \bibinfo {pages} {423}
  (\bibinfo {year} {1973})}\BibitemShut {NoStop}%
\bibitem [{\citenamefont {Janssen}(1976)}]{janssen_zfp1976}%
  \BibitemOpen
  \bibfield  {author} {\bibinfo {author} {\bibfnamefont {H.-K.}\ \bibnamefont
  {Janssen}},\ }\href@noop {} {\bibfield  {journal} {\bibinfo  {journal}
  {Zeitschrift f{\"u}r Physik B Condensed Matter}\ }\textbf {\bibinfo {volume}
  {23}},\ \bibinfo {pages} {377} (\bibinfo {year} {1976})}\BibitemShut
  {NoStop}%
\bibitem [{\citenamefont {Chow}\ and\ \citenamefont
  {Buice}(2015)}]{chow_buice_jmathneuro2015}%
  \BibitemOpen
  \bibfield  {author} {\bibinfo {author} {\bibfnamefont {C.}~\bibnamefont
  {Chow}}\ and\ \bibinfo {author} {\bibfnamefont {M.}~\bibnamefont {Buice}},\
  }\href {\doibase 10.1186/s13408-015-0018-5} {\bibfield  {journal} {\bibinfo
  {journal} {J. Math. Neurosci.}\ }\textbf {\bibinfo {volume} {5}},\ \bibinfo
  {pages} {8} (\bibinfo {year} {2015})}\BibitemShut {NoStop}%
\bibitem [{\citenamefont {Cugliandolo}(2013)}]{cugliandolo_lectures2013}%
  \BibitemOpen
  \bibfield  {author} {\bibinfo {author} {\bibfnamefont {L.~F.}\ \bibnamefont
  {Cugliandolo}},\ }\href
  {www.lpthe.jussieu.fr/~leticia/TEACHING/cours-martine-london.pdf} {\enquote
  {\bibinfo {title} {Out of equilibrium dynamics of complex systems},}\ }
  (\bibinfo {year} {2013}),\ \bibinfo {note} {lecture Notes}\BibitemShut
  {NoStop}%
\bibitem [{\citenamefont {{De Dominicis}}(1978)}]{dedominicis_prb1978}%
  \BibitemOpen
  \bibfield  {author} {\bibinfo {author} {\bibfnamefont {C.}~\bibnamefont {{De
  Dominicis}}},\ }\href {\doibase 10.1103/PhysRevB.18.4913} {\bibfield
  {journal} {\bibinfo  {journal} {Phys. Rev. B}\ }\textbf {\bibinfo {volume}
  {18}},\ \bibinfo {pages} {4913} (\bibinfo {year} {1978})}\BibitemShut
  {NoStop}%
\end{thebibliography}%
\end{document}